\documentclass[reprint,amsmath,amssymb,aps]{revtex4-1}
\usepackage{amsmath}
\usepackage{hyperref}
\usepackage{graphicx}
\usepackage{bm}
\usepackage[usenames, dvipsnames]{xcolor}

\begin{document}

\title{Proton elastic scattering on calcium isotopes from chiral nuclear optical potentials}
\author{T. R. Whitehead}
\author{Y. Lim}
\author{J. W. Holt}
\affiliation{Cyclotron Institute, Texas A\&M University, College Station, TX 77843, USA}
\affiliation{Department of Physics and Astronomy, Texas A\&M University, College Station, TX 77843, USA}

\begin{abstract}
We formulate microscopic optical potentials for nucleon-nucleus scattering from chiral two- and three-nucleon forces. The real and imaginary central terms of the optical potentials are obtained from the nucleon self energy in infinite nuclear matter at a given density and isospin asymmetry, calculated self-consistently to second order in many-body perturbation theory. The real spin-orbit term is extracted from 
the same chiral potential using an improved density matrix expansion. The density-dependent optical potential is then folded with the nuclear density distributions of $^{40,42,44,48}$Ca from which we study proton-nucleus elastic scattering and total reaction cross sections using the reaction code TALYS. We compare the results of the microscopic calculations to those of phenomenological models and experimental data up to projectile energies of $E = 180$ MeV. While overall satisfactory agreement with the available experimental data is obtained, we find that the elastic scattering and total reaction cross sections can be significantly improved with a weaker imaginary optical potential, particularly for larger projectile energies.
\end{abstract}

\maketitle

\section{Introduction}
Nucleon-nucleus optical potentials are a valuable tool for predicting a wide range of scattering and reaction processes by replacing the complicated many-body dynamics of nucleons interacting through two- and three-body forces with an average complex, energy-dependent single-particle potential. Global phenomenological optical potentials \cite{Varner91,KD03} have been constructed by fitting to experimental data spanning a large range of projectile energies across many nuclei. Although phenomenological optical potentials are very successful at describing scattering processes involving nuclei near stability, microscopic optical potentials are not tuned to experimental data and therefore may have greater predictive power for reactions involving exotic isotopes. 

Semi-microscopic global optical potentials \cite{JEUKENNE1976,LEJEUNE78} derived in the 1970's from high-precision one-boson-exchange nucleon-nucleon interactions are still widely used today \cite{Hansen85,Petler85,DelarocheILDA,bauge01,goriely07}. Whereas the density and isospin-asymmetry dependence is computed microscopically, the overall strengths of the real and imaginary volume terms are often adjusted with energy-dependent empirical strength factors. The effects of three-body forces were generally neglected in these early works, and within the original nuclear matter approach no spin-orbit optical potential could be derived. More recently, three-body forces have been implemented \cite{Holt13omp,Toyokawa3N,Holt15omp} in calculations of the real and imaginary central terms, while phenomenological spin-orbit optical potentials have been added \cite{camarda89,DelarocheILDA,bauge01} in order to better describe analyzing powers and differential cross sections at large scattering angles. An alternative approach \cite{kerman59,BRIEVA1977299,Hoffmann85,Elster90,Arellano90,arellano95,vorabbi16,Vorabbi18} to constructing microscopic optical potentials is based on multiple scattering theory involving the nucleon-nucleon $T$-matrix. Such an approach naturally generates a spin-orbit contribution, but the implementation of medium effects \cite{chin93,crespo93} and three-body forces remains challenging, which in practice often limits the theory to large scattering energies $E\gtrsim 200$\,MeV. Other current approaches to deriving predictive optical potentials include the self-consistent Green's function method \cite{waldecker11} and the dispersive optical model \cite{Dickhoff07}.

Recently there has been much interest in the development of microscopic optical potentials \cite{Holt13omp,Egashira,Holt15omp,Rotureau18,Rotureau17,Vorabbi18,Toyokawa15,DURANT18} based on chiral effective field theory (EFT) \cite{WEINBERG79,epelbaum09,MACHLEIDT11}. The main motivation is to implement more realistic microphysics involving multi-pion exchange contributions to the nuclear force, three-body interactions, and theoretical uncertainty estimates. Chiral optical potentials are well suited to describe low-energy scattering processes but are expected to break down for energies approaching the relevant momentum-space cutoff employed. In practice, the presence of the cutoff constrains nucleon projectile energies to lie below $E \lesssim 200$\,MeV.

In the present study, we aim to lay the groundwork for a revised nuclear matter description of the global nucleon-nucleus optical potential based on chiral EFT. Ultimately the goal will be to develop a theory for nucleon-nucleus scattering across a large range of isotopes, including those off stability, at energies up to 200 MeV. As a starting point we consider differential elastic and total reaction cross sections for proton-nucleus scattering along a chain of calcium isotopes, $^{40,42,44,48}$Ca, at energies ranging from 2-160 MeV where experimental data are available. We also compare to the global phenomenological optical potential of Koning and Delaroche \cite{KD03} and investigate to what extent modern phenomenological parametrizations of the optical potential are consistent with microscopic analyses. Our calculations are performed within the TALYS \cite{TALYS} reaction code for which we have developed an implementation of our microscopic optical potential.

We take as a starting point for the calculation a particular high-precision 2N + 3N chiral nuclear potential with momentum-space cutoff $\Lambda = 450$\,MeV. The low-energy constants of the potential are fitted to nucleon-nucleon scattering phase shifts, deuteron properties, and in the case of three-body contact terms also the triton binding energy and lifetime. The nucleon-nucleon interaction is taken at next-to-next-to-next-to-leading order (N3LO), while only the N2LO three-nucleon force is included. The inconsistent treatment of two- and three-body forces at the level of the chiral expansion is undesirable, but work toward fully consistent two- and many-body forces is in progress \cite{tews13,drischler16,kaiser18}. We note that the chiral nuclear potential employed in the present work exhibits good nuclear matter properties (saturation energy and density \cite{coraggio14}, thermodynamics \cite{wellenhofer14,wellenhofer15}, and Fermi liquid parameters \cite{Holt18}) when calculated at least to second order in many-body perturbation theory. In the future we plan to perform calculations of the nucleon-nucleus optical potential from a wider range of high-precision chiral nuclear forces in order to better assess theoretical uncertainties.

In quantum many-body theory, the optical potential for scattering states is identified with the energy- and momentum-dependent single-particle self energy \cite{PhysRevLett.3.96}. We first compute the nucleon self energy in homogeneous nuclear matter at arbitrary density and composition (proton fraction) from chiral two- and three-body forces to second order in many-body perturbation theory. We next compute nuclear density distributions for selected calcium isotopes ($^{40}$Ca, $^{42}$Ca, $^{44}$Ca, $^{48}$Ca) from mean field theory employing recently derived \cite{Lim17} Skyrme effective interactions constrained by chiral effective field theory. In the local density approximation (LDA) the nucleon-nucleus optical potential is computed \cite{Jeukenne77lda} by folding the nucleon self-energy in homogeneous matter with the derived density distributions. Since the LDA is known \cite{Jeukenne77lda} to underestimate the surface diffuseness of the optical potential in finite nuclei, we employ the improved local density approximation (ILDA) described in Refs.\ \cite{Jeukenne77lda,DelarocheILDA} to account for the non-zero range of the nuclear force. 

The method outlined thus far is versatile since it can be used to produce optical potentials for a very wide range of nuclei. However, the LDA nuclear matter approach cannot capture the physics of collective surface modes, shell effects \cite{Kosugi82}, and surface-peaked spin-orbit optical potentials. The latter are particularly important for spin observables and elastic scattering cross sections at large angles. In the present work we therefore construct a spin-orbit optical potential from the improved density matrix expansion \cite{bogner08,gebremariam10,KaiserHoltEDF}, which improves the description of the spin-dependent part of the energy density functional compared to the standard density matrix expansion of Negele and Vautherin \cite{negele72}. We then benchmark our approach to experimental data for proton elastic scattering and total reaction cross sections on the calcium isotopes $^{40}$Ca, $^{42}$Ca, $^{44}$Ca, $^{48}$Ca.

The paper is organized as follows. In Section \ref{chiral} we describe details of the microscopic calculation of the optical potential in nuclear matter from chiral EFT. We then calculate nuclear density distributions from mean field theory and employ the ILDA to construct nucleon-nucleus optical potentials for calcium isotopes. The microscopic optical potentials are parametrized in the form of the Koning-Delaroche (KD) \cite{KD03} phenomenological optical potential in order to implement them into the reaction code TALYS. In Section \ref{results} we compute proton-nucleus elastic differential scattering cross sections up to projectile energy $E = 160$\,MeV and total reaction cross sections up to $E=180$\,MeV. These results are compared to empirical data and predictions from the KD phenomenological optical potential. We end with a summary and conclusions.


\section{Optical potential from chiral effective field theory}
\label{chiral}

\subsection{Real and imaginary central terms}
\begin{figure}[t]
\includegraphics[scale=0.45]{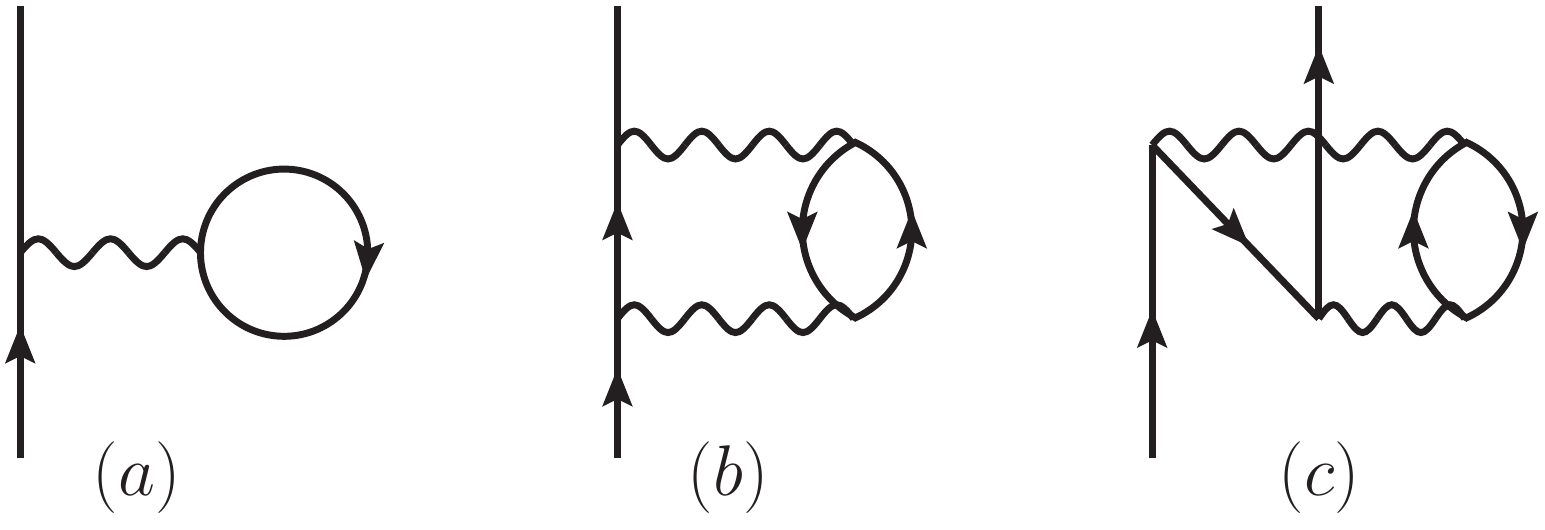}
\caption{Diagrammatic representations of the first and second order contributions to the self energy. The solid lines represent nucleon propagators and the wavy lines represent the in medium two-nucleon interaction.}
\label{fig:diagrams}
\end{figure}
In recent work \cite{Holt13omp,Holt15omp} the nucleon self energy in homogeneous nuclear matter has been computed employing a set of nuclear potentials derived from chiral effective field theory. The first- and second-order perturbative contributions to the nucleon self energy are shown graphically in Fig.\ \ref{fig:diagrams} and given quantitatively by
\begin{equation}
\Sigma^{(1)}_{2N}(q;k_f)=\sum_{1} \langle \vec{q} \: \vec{h}_1 s s_1 t t_1 | \bar{V}_{2N}^{\rm eff} | \vec{q} \: \vec{h}_1 s s_1 t t_1 \rangle n_1 ,
\label{eq:2}
\end{equation}
\begin{eqnarray}
\label{sig2a}
&&\hspace{-.1in} \Sigma^{(2a)}_{2N}(q,\omega;k_f) \\ \nonumber
&&= \frac{1}{2} \sum_{123} \frac{|\langle \vec{p}_1 \vec{p}_3 s_1 s_3 t_1 t_3 | \bar{V}_{2N}^{\rm eff} | \vec{q} \vec{h}_2 s s_2 t t_2 \rangle|^2}{\omega+\epsilon_2-\epsilon_1-\epsilon_3+i\eta} \bar{n}_1 n_2 \bar{n}_3,
\end{eqnarray}
\begin{eqnarray}
\label{sig2b}
&&\hspace{-.1in} \Sigma^{(2b)}_{2N}(q,\omega;k_f) \\ \nonumber
&&= \frac{1}{2} \sum_{123} \frac{|\langle \vec{h}_1 \vec{h}_3 s_1 s_3 t_1 t_3 | \bar{V}_{2N}^{\rm eff} | \vec{q} \vec{p}_2 s s_2 t t_2 \rangle|^2}{\omega+\epsilon_2-\epsilon_1-\epsilon_3-i\eta}   n_1 \bar{n}_2 n_3,
\end{eqnarray}
where $n_i$ is the occupation probability $\theta(k_f - k_i)$ for a filled state with momentum $\vec k_i$ below the Fermi surface, the occupation probability for particle states is $\bar n_i = \theta(k_i-k_f)$, the summation is over intermediate-state momenta for particles $\vec p_i$ and holes $\vec h_i$, their spins $s_i$, and isospins $t_i$. The nuclear potential $\bar V_{2N}^{\rm eff}$ represents the antisymmetrized two-body interaction consisting of the bare nucleon-nucleon (NN) potential $V_{NN}$ together with an effective, medium-dependent NN interaction $V_{NN}^{\rm med}$ derived from the N2LO chiral three-nucleon force by averaging one particle over the filled Fermi sea of noninteracting nucleons \cite{bogner05,Holt09,hebeler10}. In the first-order Hartree-Fock contribution, Eq.\ (\ref{eq:2}), the effective interaction is given by $\bar V_{2N}^{\rm eff} = V_{NN} + \frac{1}{2}V_{NN}^{\rm med}$, while for the higher-order contributions, Eqs.\ (\ref{sig2a}) and (\ref{sig2b}), the effective interaction is given by $\bar V_{2N}^{\rm eff} = V_{NN} + V_{NN}^{\rm med}$. The Hartree-Fock contribution is nonlocal, energy-independent, and purely real, while the second-order contributions are in general nonlocal, energy-dependent, and complex. The single-particle energies in the denominators of Eqs.\ (\ref{sig2a}) and (\ref{sig2b}) are computed self-consistently according to
\begin{equation}
\label{eq:1}
\epsilon(q)=\frac{q^2}{2M}+\text{Re} \Sigma(q,\epsilon(q)),
\end{equation}
where $M$ is the free-space nucleon mass.

In the present work the self energy is computed for arbitrary isospin-asymmetry, $\delta_{np} = (\rho_n-\rho_p)/(\rho_n+\rho_p)$, which is essential for an accurate description of nuclei for which $N\ne Z$. The resulting optical potentials for nucleons propagating in homogeneous matter characterized by its proton and neutron Fermi momenta $k_f^p$ and $k_f^n$ are given by
\begin{eqnarray}
U_p(E;k_f^p,k_f^n) &=& V_p(E;k_f^p,k_f^n) + i W_p(E;k_f^p,k_f^n), \nonumber \\ 
U_n(E;k_f^p,k_f^n) &=& V_n(E;k_f^p,k_f^n) + i W_n(E;k_f^p,k_f^n)
\label{omp}
\end{eqnarray}
with
\begin{equation}
V_i(E;k_f^p,k_f^n) = {\rm Re}\Sigma_i(q,E(q);k_f^p,k_f^n),
\end{equation}
\begin{equation}
W_i(E;k_f^p,k_f^n) = \frac{M_i^{k*}}{M} {\rm Im}\Sigma_i(q,E(q);k_f^p,k_f^n),
\end{equation}
where the subscript $i$ denotes a propagating proton or neutron. In relating the physical imaginary part of the optical potential to the imaginary part of the nucleon self-energy we have multiplied \cite{Negele81,fantoni81} by the effective $k$-mass $M_i^{k*}$ defined by
\begin{equation}
\frac{M_i^{k*}}{M} = \left ( 1 + \frac{M}{k}\frac{\partial}{\partial k}V_i(k,E(k) \right )^{-1},
\end{equation}
in order to account for the non-locality of the optical potential.

\subsection{Spin-orbit optical potential}
The effective one-body spin-orbit interaction vanishes in homogeneous nuclear matter and therefore cannot be computed within the framework described above. Instead we employ an improved density matrix expansion \cite{gebremariam10,Gebremariam10npa,KaiserHoltEDF} to construct the one-body spin-orbit interaction from chiral two- and three-body forces. The improved density matrix expansion takes advantage of phase space averaging to derive a more accurate spin-dependent energy density functional compared to the standard density matrix expansion of Negele-Vautherin \cite{negele72}. 

From the definition of the density matrix 
\begin{equation}
\rho(\vec r_1 \sigma_1 \tau_1; \vec r_2 \sigma_2 \tau_2) = \sum_\alpha \Psi_\alpha^*(\vec r_2 \sigma_2 \tau_2)
\Psi_\alpha(\vec r_1 \sigma_1 \tau_1),
\end{equation}
where $\Psi_\alpha$ are energy eigenfunctions associated with occupied orbitals of the non-relativistic many-body system, the energy density functional for $N=Z$ even-even nuclei in the Hartree-Fock approximation can be expanded up to second order in spatial gradients as
\begin{eqnarray}
{\cal E}[\rho,\tau,\vec J\,] &=& \rho\,\bar E(\rho)+\bigg[\tau-
{3\over 5} \rho k_f^2\bigg] \bigg[{1\over 2M_N}+F_\tau(\rho)
\bigg] \\  \nonumber && + (\vec \nabla \rho)^2\, F_\nabla(\rho)+  \vec \nabla
\rho \cdot\vec J\, F_{SO}(\rho)+ \vec J\,^2 \, F_J(\rho)\, ,
\label{edf}
\end{eqnarray}
where $\rho(\vec r\,) =2k_f^3(\vec r\,)/3\pi^2 =  \sum_\alpha \Psi^\dagger_\alpha(\vec r\,) \Psi_\alpha(\vec r\,)$ defines the local density with $k_f(\vec r\,)$ the local Fermi momentum, $\tau(\vec r\,) =  \sum_\alpha \vec \nabla \Psi^\dagger_\alpha (\vec r\,) \cdot \vec \nabla \Psi_\alpha(\vec r\,)$ is the kinetic energy density, and $\vec J(\vec r\,) = i \sum_\alpha \vec \Psi^\dagger_\alpha(\vec r\,) \vec \sigma \times \vec \nabla \Psi_\alpha(\vec r\,)$ is the spin-orbit density. These terms are multiplied by the density-dependent strength functions $\bar E(\rho), F_\tau(\rho), F_\nabla(\rho), F_{SO}(\rho), F_J(\rho)$, of which we are presently only interested in the spin-orbit term $F_{SO}(\rho)$. In effect, the spin-orbit optical potential is therefore calculated for $N=Z$ nuclei to first order in many-body perturbation theory. In the future, higher-order perturbative contributions \cite{zhang18} to the microscopic nuclear energy density functional may be investigated. We note that we do not include the isovector part \cite{Kaiser12} of the spin-orbit interaction for $N \neq Z$ nuclei in this study since it is known to be small compared to the isoscalar part \cite{KaiserHoltEDF}. 

\subsection{Improved local density approximation}

\begin{figure}[t]
\includegraphics[scale=0.47]{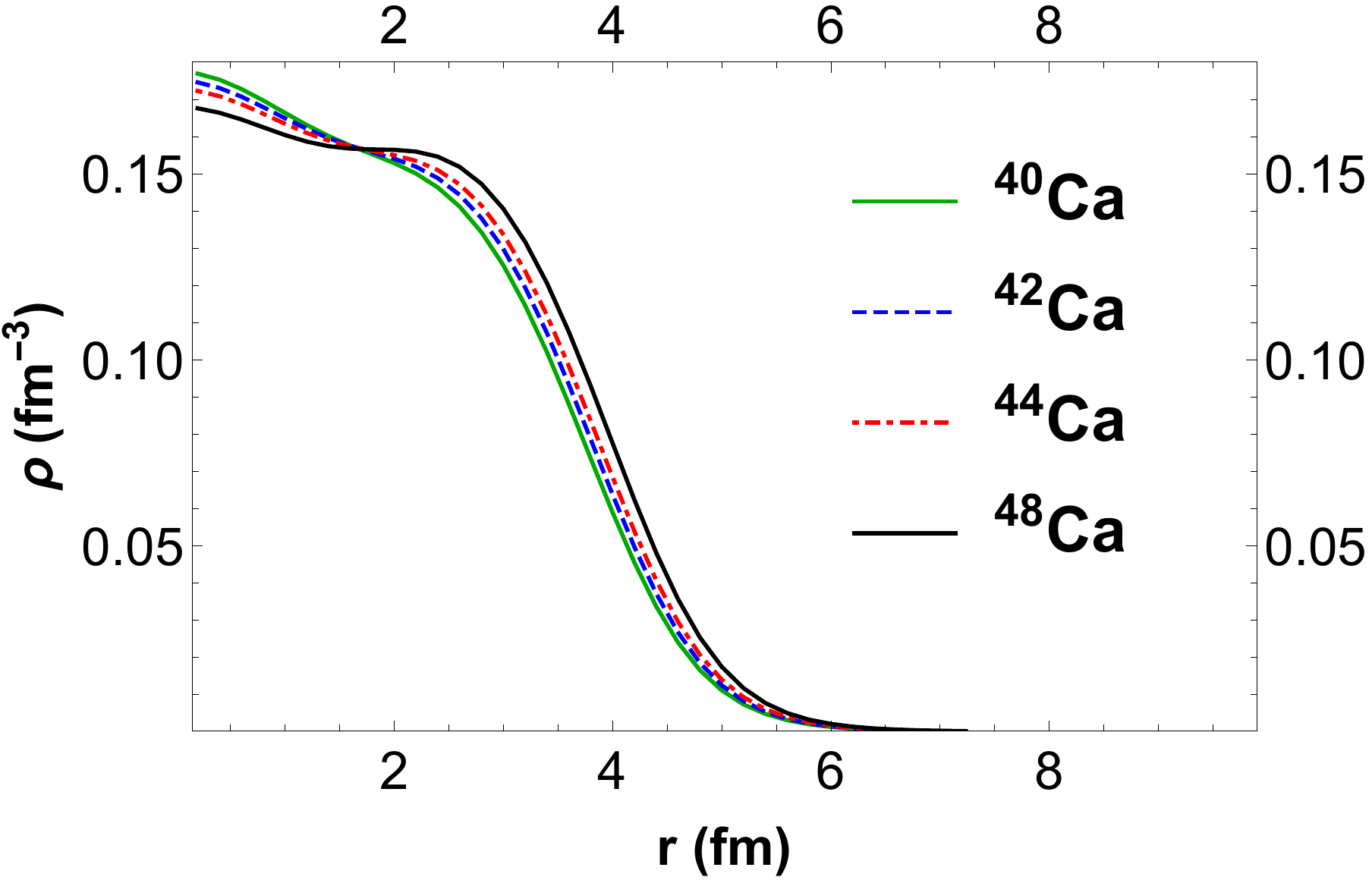}
\caption{The nucleon density distributions for $^{40,42,44,48}$Ca calculated in mean field theory from the Skyrme Sk$\chi$450 effective interaction constrained by chiral effective field theory.}
\label{fig:Ca40denprof}
\end{figure}

\begin{figure*}[t]
	\includegraphics[scale=0.38]{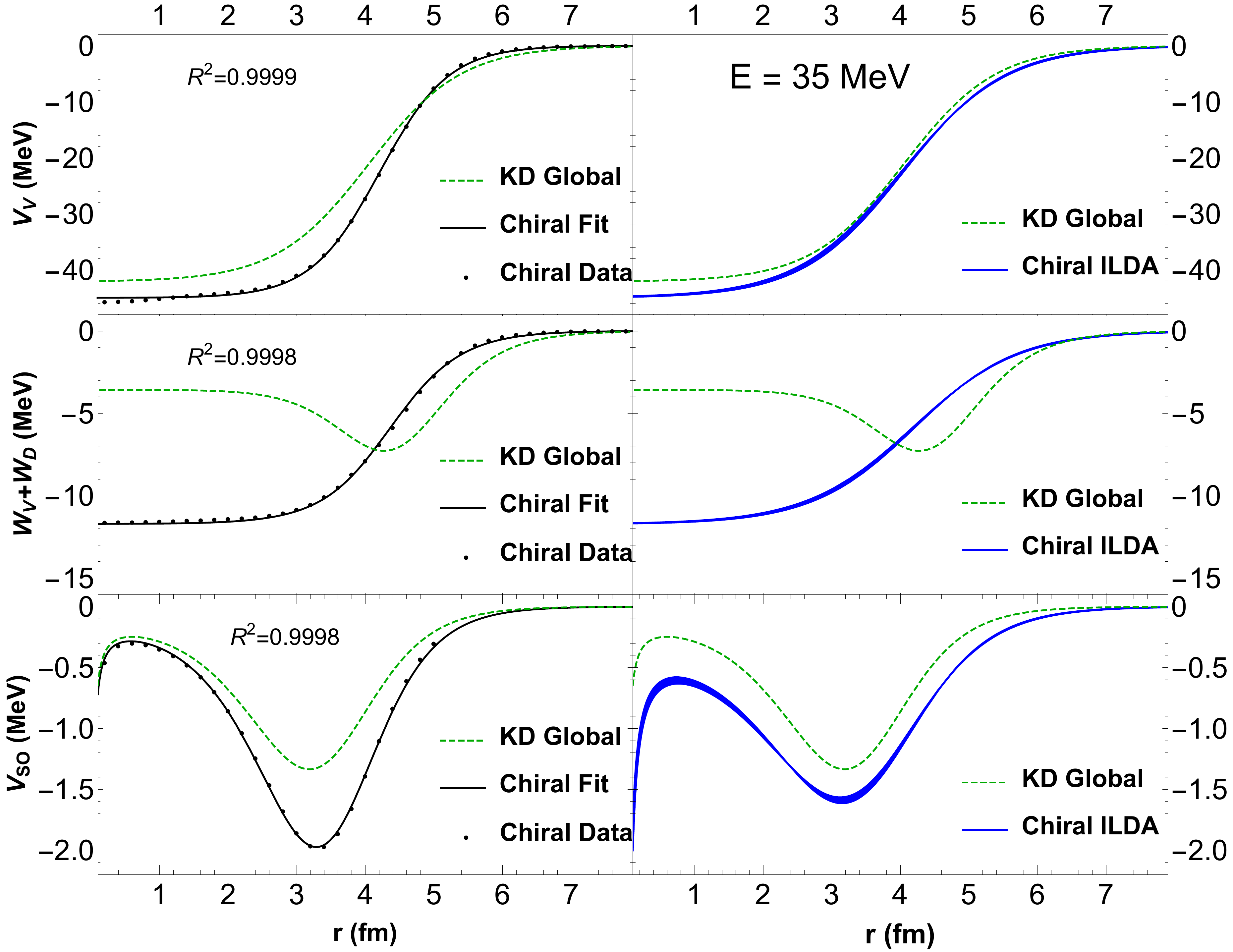}
	\caption{Left (right): The real, imaginary, and spin-orbit terms of the microscopic optical potential for proton-$^{40}$Ca scattering at $E=2.35$\,MeV before (after) applying the improved local density approximation. In the left panel the dots represent raw results from chiral EFT, while the solid black lines represent fits to the Koning-Delaroche (KD) form. We also compare to the global KD phenomenological optical potential, shown as the dashed green line, at the same energy.}
	\label{fig:potentialplots2}
\end{figure*}

We employ the improved local density approximation (ILDA) to construct the nucleon-nucleus optical potential for finite nuclei. The density dependent optical potential (both central and spin-orbit parts) is folded with the radial density distribution of a target nucleus. The nuclear density distributions are calculated within mean field theory from the Sk$\chi$450 Skyrme interaction \cite{Lim17}, which fits both finite nuclei properties as well as theoretical calculations of the asymmetric nuclear matter equation of state from the N3LO $\Lambda = 450$\,MeV chiral potential used in the calculation of the self energy. In Fig.\ \ref{fig:Ca40denprof} we show the resulting nucleon density distributions for each of the calcium isotopes $^{40}$Ca, $^{42}$Ca, $^{44}$Ca, $^{48}$Ca.

In the standard local density approximation, the strength of the nucleon-nucleus optical potential at a given radial distance $r$ is evaluated as
\begin{eqnarray}
&&\hspace{-.2in}V(E;r) + i W(E;r) = V(E;k_f^p(r),k_f^n(r)) \\ \nonumber
&& + i W(E;k_f^p(r),k_f^n(r)),
\end{eqnarray}
where $k_f^p(r)$ and $k_f^n(r)$ are the local proton and neutron Fermi momenta.
This approximation is strictly valid only for zero-range nuclear forces, and when applied to nucleon-nucleus optical potentials it is known to underestimate the surface diffuseness \cite{BRIEVA1977317,Jeukenne77lda}. Consequently, such an approach is inadequate for an accurate description of nuclear elastic scattering and reaction processes. The improved local density approximation applies a Gaussian smearing
\begin{equation}\label{eq:ilda}
{V}(E;r)_{ILDA}=\frac{1}{(t\sqrt{\pi})^3}\int V(E;r') e^{\frac{-|\vec{r}-\vec{r}'|^2}{t^2}} d^3r'
\end{equation}
characterized by an adjustable length scale $t$ associated with the non-zero range of the nuclear force. In the limiting case of $t \rightarrow 0$, a factor of $\delta(|\vec{r}-\vec{r'}|)$ replaces the Gaussian, giving ${V}_{ILDA}(E;r) \rightarrow V(E;r)$. In Ref. \cite{DelarocheILDA} it is found that for the central part of the interaction $t_C=1.2\,{\rm fm}$ gives the best fit to experimental reaction cross sections for $10\,{\rm MeV} < E < 200 \,{\rm MeV}$ and targets ranging from $^{40}$Ca to $^{208}$Pb. In the present work we vary the range parameter $1.15\,{\rm fm} < t_C < 1.25\,{\rm fm}$. This variation is used to estimate the theoretical uncertainty associated with our choice of the length scale $t_C$. For the spin-orbit part of the optical potential, we estimate the range parameter $t_{SO}$ from the root mean square radius of the Argonne $v_{18}$ spin-orbit nucleon-nucleon potential \cite{Argonne}. We found $t_{SO}=1.07$ fm and took two values, $t _{SO}= 1.0 , 1.1$\,fm to estimate the uncertainty. 

\begin{figure*}[t]
	\includegraphics[scale=0.38]{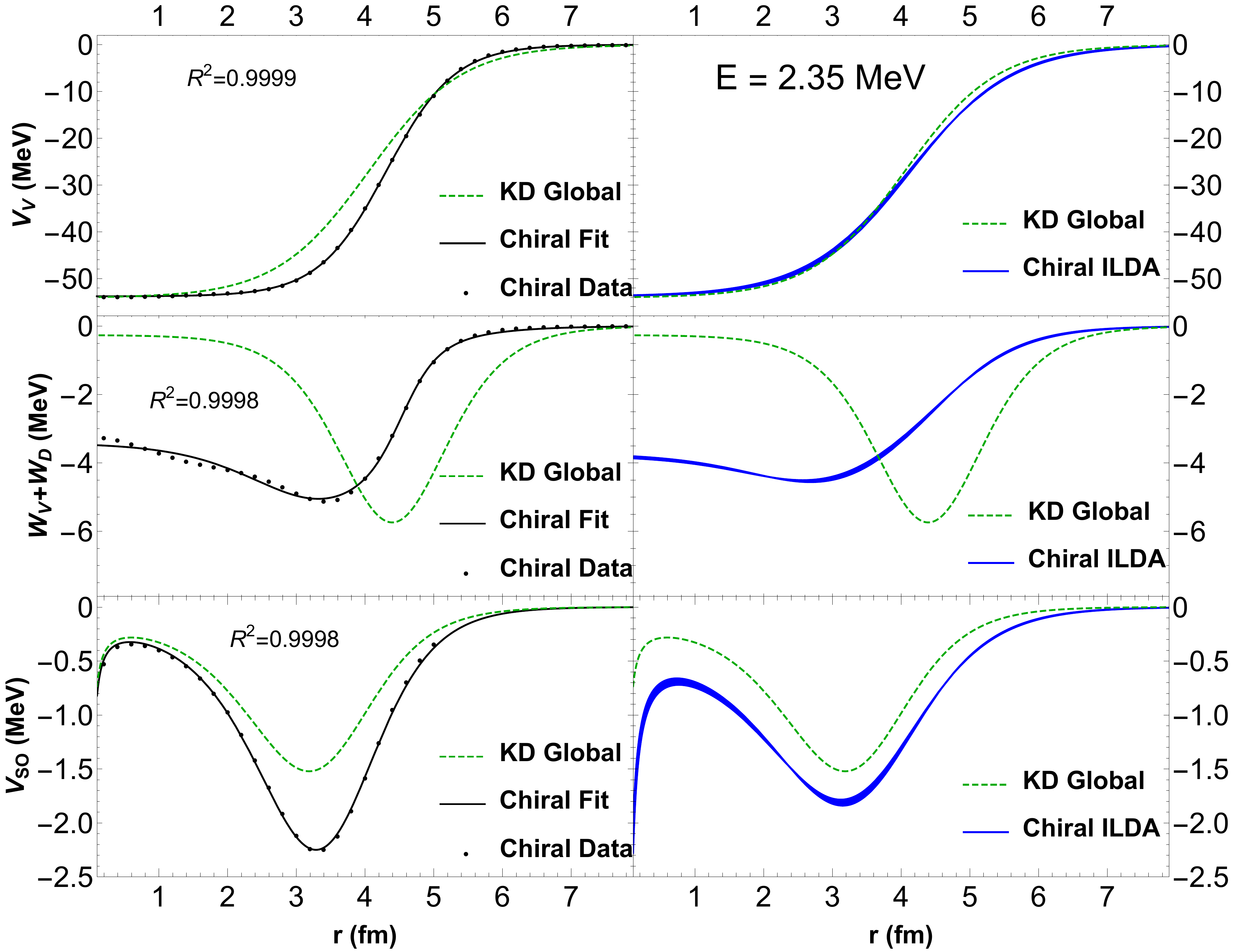}
	\caption{Same as Fig.\ \ref{fig:potentialplots2}, except that the projectile energy is $E=35$\,MeV.}
	\label{fig:potentialplots35}
\end{figure*}

In Figs.\ \ref{fig:potentialplots2} and \ref{fig:potentialplots35} we show the comparison between the central and spin-orbit optical potentials in the LDA (left panels) and ILDA (right panels) for the proton-$^{40}$Ca optical potential at projectile energies $E=2.35$\,MeV and $E=35$\,MeV respectively. In the left panels, the dots indicate the results from chiral effective field theory, and for comparison we show as the green dashed lines the phenomenological optical potentials from Koning and Delaroche \cite{KD03}. In particular, the real central part of the microscopic optical potential has a much smaller surface diffuseness compared to phenomenology. When the improved local density approximation is employed in the right panels (solid blue bands), the comparison to phenomenology is much improved. In addition, the overall strength of the real central part of the optical potential is in very good agreement with the Koning-Delaroche phenomenological optical potential. Varying the ILDA range parameter $t_C$ between $1.15\,{\rm fm}$ and $1.25\,{\rm fm}$ yields only a small change in the overall shape of the real central part, which suggests that the theoretical predictions for scattering cross sections computed in the next section will not be especially sensitive to the precise choice of $t_C$.

In the middle panels of Figs.\ \ref{fig:potentialplots2} and \ref{fig:potentialplots35} we plot the imaginary part of the microscopic optical potential for proton-$^{40}$Ca scattering at $E=2.35$\,MeV and $E=35$\,MeV. This is again compared to the phenomenological optical potential of Koning and Delaroche, which is written as the sum of a volume imaginary part $W_V$ and a surface imaginary part $W_D$. In the microscopic description of the imaginary part, there is no distinction between these two components. We observe that in contrast to the real central part of the optical potential, the microscopic imaginary part exhibits large qualitative differences compared to phenomenology. The most striking difference is a much smaller surface peak, which only appears at low projectile energies in the microscopic calculation but persists to much higher energies ($E \simeq 100$\,MeV) in the phenomenological optical potential. For instance, at the scattering energy $E=35$\,MeV, the surface peak has essentially vanished in the microscopic calculation, and the remaining ``volume'' imaginary part is large compared to phenomenology. In fact, this is a common feature \cite{LEJEUNE78,Lagrange82,Kohno84,Petler85} in microscopic optical potentials computed in the nuclear matter approach. Modern semi-microscopic optical potentials therefore include empirical energy-dependent strength factors multiplying the real and imaginary central parts \cite{DelarocheILDA,goriely07}.

In the bottom panels of Figs.\ \ref{fig:potentialplots2} and \ref{fig:potentialplots35} we show the real spin-orbit part of the microscopic optical potential compared to the Koning-Delaroche phenomenological optical potential for proton-$^{40}$Ca scattering at $E=2.35$\,MeV and $E=35$\,MeV. The radial shape of the microscopic spin-orbit optical potential is found to be very similar to that of the Koning-Delaroche optical potential, however, the strength of the microscopic potential is larger. Indeed, the density matrix expansion carried out at the Hartree-Fock level is known \cite{KaiserHoltEDF,holt16pr} to produce a stronger spin-orbit interaction, by about 20-50\%, than is required from traditional mean field theory studies of finite nuclei. Higher-order perturbative contributions are expected to remedy this feature. In particular, multi-pion-exchange processes have been shown \cite{kaiser10} to reduce the strength of the one-body spin-orbit interaction in finite nuclei. This provides additional motivation for including $G$-matrix correlations in the density matrix expansion as outlined in \cite{zhang18}. As in the case of the central components of the optical potential, we find relatively small differences between spin-orbit potentials produced with two choices of the ILDA length scale $t _{SO}=1.0,1.1$\,fm.

\begingroup
\renewcommand{\arraystretch}{1.25}
\setlength{\tabcolsep}{15pt}
\begin{table}[t]
\caption{Shape parameters for the proton-$^{40}$Ca microscopic optical potential at the three energies $E=2.35,35,100$\,MeV. Also shown are the corresponding shape parameters for the Koning-Delaroche (KD) optical potential that are independent of energy.} 
\centering
\begin{tabular}{|c | c| c | c |} 
\hline
$E=2.35$ MeV & $V_V$& $W_V$ & $V_{SO}$ \\ 
\hline 
$r$ (fm)& 1.213 & 1.344 & 1.015 \\
$a$ (fm)& 0.723 & 0.583 & 0.706 \\
\hline
$E=35$ MeV & $V_V$& $W_V$ & $V_{SO}$ \\ 
\hline 
$r$ (fm)& 1.183 & 1.204 & 1.015 \\
$a$ (fm)& 0.730 & 0.751 & 0.706 \\
\hline 
$E=100$ MeV & $V_V$& $W_V$ & $V_{SO}$ \\ 
\hline 
$r$ (fm)& 1.173 & 0.846 & 1.015 \\
$a$ (fm)& 0.713 & 0.702 & 0.706 \\
\hline
KD & $V_V$& $W_V$ & $V_{SO}$ \\ 
\hline 
$r$ (fm)& 1.185 & 1.185 & 0.996\\
$a$ (fm)& 0.672 & 0.672 & 0.590\\
\hline 
\end{tabular}
\label{table:ShapeParameters} 
\end{table}
\endgroup

\subsection{Parameterization of the chiral optical potential}
In order to facilitate the implementation of our microscopic optical potential into standard nuclear reaction codes, such as TALYS, we fit our optical potential to the phenomenological form of Koning and Delaroche. Eventually our aim is to construct a global microscopic optical potential and make it available in a convenient form for nuclear reaction practitioners. This exercise may also help to reveal any deficiencies in the assumed form of phenomenological optical potentials. We recall that in the phenomenological description, the optical potential takes the form
\begin{eqnarray}
&&\hspace{-.3in} U(r,E) = V_V(r,E) + i W_V(r,E) + i W_D(r,E) \label{phen} \\ \nonumber 
&& + V_{SO}(r,E) \vec \ell \cdot \vec s + i W_{SO}(r,E) \vec \ell \cdot \vec s + V_C(r),
\end{eqnarray}
consisting of a real volume term, an imaginary volume and surface term, a real and imaginary spin-orbit term, and finally the central Coulomb interaction. In Eq.\ (\ref{phen}), $\vec \ell$ and $\vec s$ are the single-particle orbital angular momentum and spin angular momentum operators, respectively. Since the phenomenological imaginary spin-orbit term is very small and cannot be extracted within the present microscopic approach, we neglect it in the rest of the discussion. The energy and radial dependence of the different terms in the phenomenological optical potential are assumed to factorize according to
\begin{equation}
V_V(r,E) = {\cal V}_V(E) f(r;r_V,a_V),
\label{vreal}
\end{equation}
\begin{equation}
W_V(r,E) = {\cal W}_V(E) f(r;r_W,a_W),
\label{vim}
\end{equation}
\begin{equation}
W_D(r,E) = -4a_D{\cal W}_D(E) \frac{d}{dr} f(r;r_D,a_D),
\label{sim}
\end{equation}
\begin{equation}
V_{SO}(r,E) = {\cal V}_{SO}(E) \frac{1}{m_\pi^2}\frac{1}{r}\frac{d}{dr} f(r;r_{SO},a_{SO}),
\label{soreal}
\end{equation}
where
\begin{equation}
f(r;r_i,a_i) = \frac{1}{1+e^{(r-A^{1/3}r_i)/a_i}}
\end{equation}
is of the Woods-Saxon form with $A$ the mass number and \{$r_i$,$a_i$\} the energy-independent geometry parameters that encode the size and diffuseness of a given target nucleus respectively. In phenomenological optical potentials, these shape parameters vary weakly with the target nucleus. The energy-dependent strength functions in the KD parametrization have the form
\begin{equation}
{\cal V}_V(E)= v_1 ( 1 - v_2 \tilde{E} + v_3 \tilde{E}^2 - v_4 \tilde{E}^3),
\label{vreale}
\end{equation}
\begin{equation}
{\cal W}_V(E)= w_1 \frac{ \tilde{E}^2 }{ \tilde{E}^2 + w^{2}_{2} },
\label{vime}
\end{equation}
\begin{equation}
{\cal W}_D(E)= d_1 \frac{\tilde{E}^2 \: e^{-d_2\tilde{E}}}{\tilde{E}^2+d^{2}_{3}},
\label{sime}
\end{equation}
\begin{equation}
{\cal V}_{SO}(E)= v_{SO1} e^{-v_{SO2} \tilde{E}},
\label{soreale}
\end{equation}
where $\tilde{E} = E - E_F$ is the projectile energy relative to the Fermi energy $E_F$. 

In the left panels of Figs.\ \ref{fig:potentialplots2} and \ref{fig:potentialplots35} we show as the solid black lines the best fit functions of the form Eqs.\ (\ref{vreal})-(\ref{soreale}) to the microscopic calculations. We see that overall the phenomenological form can reproduce well the radial dependence of the microscopic optical potential. In the present study we have isolated the optical potential at low energy, where there is a defined surface imaginary peak, and fitted to the phenomenological form separately. At larger energies $E>50$\,MeV, we have also fitted to the phenomenological form separately. To show that no crucial features of the chiral potential are lost in this parameterization, we also display in Figs.\ \ref{fig:potentialplots2} and \ref{fig:potentialplots35} the accompanying coefficient of determination defined by
\begin{equation}
R^2 = 1 - \frac{\sum_{i} (y_i-f(r_i))^2}{\sum_{i} (y_i-\bar{y})^2},
\end{equation}
where $y_i$ represents the value of the potential from chiral EFT at location $r_i$, $f(r_i)$ is the value of the fitted function, and $\bar y$ is the mean of the chiral optical potential values. In the right panels of Figs.\ \ref{fig:potentialplots2} and \ref{fig:potentialplots35}, the ILDA results are obtained from the parametrized form of the corresponding parts of the optical potentials.

For the global KD phenomenological optical potentials we note that the real and imaginary volume terms have identical Woods-Saxon shape functions. From the microscopic perspective there is little justification for this assumption. In fact, we find that the shape parameters of the real and imaginary central optical potentials have to be fitted separately in order to achieve a good fit to the microscopic results. In Table \ref{table:ShapeParameters} we show the values of all shape parameters for the three energy windows over which we fit to the phenomenological form together with those of the phenomenological Koning-Delaroche optical potential, whose shape parameters are independent of energy. For these results we have chosen the values $t_C = 1.15$\,fm and $t_{SO} = 1.0$\,fm in the ILDA. We see that there is not a very large energy dependence in the shape parameters of the real volume part of the microscopic optical potential, but there is a significant difference among the real volume, imaginary volume, and real spin-orbit parts.

\begingroup
\renewcommand{\arraystretch}{1.25}
\setlength{\tabcolsep}{8pt}
\begin{table}[t]
\caption{Volume integrals for the real central $V_C$, imaginary central $W_C$, and real spin-orbit $V_{SO}$ parts of the proton-$^{40}$Ca optical potential. Results are shown for the microscopic chiral optical potential and for the phenomenological Koning-Delaroche (KD) optical potential.} 
\centering 
\begin{tabular}{|c | c| c |} 
\hline
$E=2.35$\,MeV & Chiral (MeV fm$^{3}$) & KD (MeV fm$^{3}$) \\ 
\hline 
$V_C$ & 524 & 480  \\ 
$W_C$ & 64 & 80 \\
$V_{SO}$ & 19 & 13 \\ 
\hline
$E=35$\,MeV & Chiral (MeV fm$^{3}$) & KD (MeV fm$^{3}$) \\ 
\hline 
$V_C$ & 413 & 374 \\ 
$W_C$ & 120 & 113 \\
$V_{SO}$ & 17 & 11 \\
\hline 
$E=100$\,MeV & Chiral (MeV fm$^{3}$) & KD (MeV fm$^{3}$) \\ 
\hline 
$V_C$ & 236 & 220  \\ 
$W_C$ & 163 & 109 \\
$V_{SO}$ & 13 & 9 \\
\hline 
\end{tabular}
\label{table:VolumeIntegrals} 
\end{table}
\endgroup

Finally, we note that the microscopic real spin-orbit optical potential calculated from the density matrix expansion has no energy dependence. The phenomenological energy dependence used in TALYS ($v_{SO2}=0.004$) is constant across all nuclei, and in fact since $v_{SO2}$ is small, the real spin-orbit term does not strongly depend on the energy. We have therefore incorporated this phenomenological energy dependence into our parametrization of the spin-orbit optical potential. As mentioned above, the imaginary spin-orbit part of the optical potential is neglected since its magnitude for the relevant energy range is {\raise.17ex\hbox{$\scriptstyle\sim$}} 0.1 MeV and has been shown to have a negligible effect on elastic scattering cross sections at relatively low energies \cite{BRIEVA1978206}.

\subsection{Volume integrals of the real and imaginary parts of the optical potential}
We end this section by comparing the volume integrals of the various components of the microscopic optical potential to those from phenomenology. It has been demonstrated \cite{JEUKENNE1976} that physical scattering observables can remain unchanged even if the various parameters of an optical potential are allowed to vary, provided that the volume integrals, defined by 
\begin{equation}
\frac{J}{A} = \frac{1}{A} \int U(r) d^{3} r,
\end{equation}
remain roughly constant. In Table \ref{table:VolumeIntegrals} we show the volume integrals for each term of the microscopic and phenomenological optical potentials for the proton-$^{40}$Ca system at the three energies $E = 2.35, 35, 100$\,MeV. We see that the microscopic real volume and spin-orbit terms are both slightly larger than their phenomenological counterparts for all three energies considered. The central imaginary term features a volume part that grows with energy and a surface peak that diminishes with energy. The imaginary term of the chiral optical potential overestimates the volume component and underestimates the surface component. For $E=35$ MeV these competing effects nearly cancel out and the chiral volume integral is close to the phenomenological volume integral. At $E=2.35$ MeV, the volume integral for the chiral imaginary term is smaller than the KD model since its surface peak is at a smaller $r$ value. For higher energies, the microscopic imaginary term becomes larger than the phenomenological imaginary term.

\begin{figure}[t]
\includegraphics[scale=0.4]{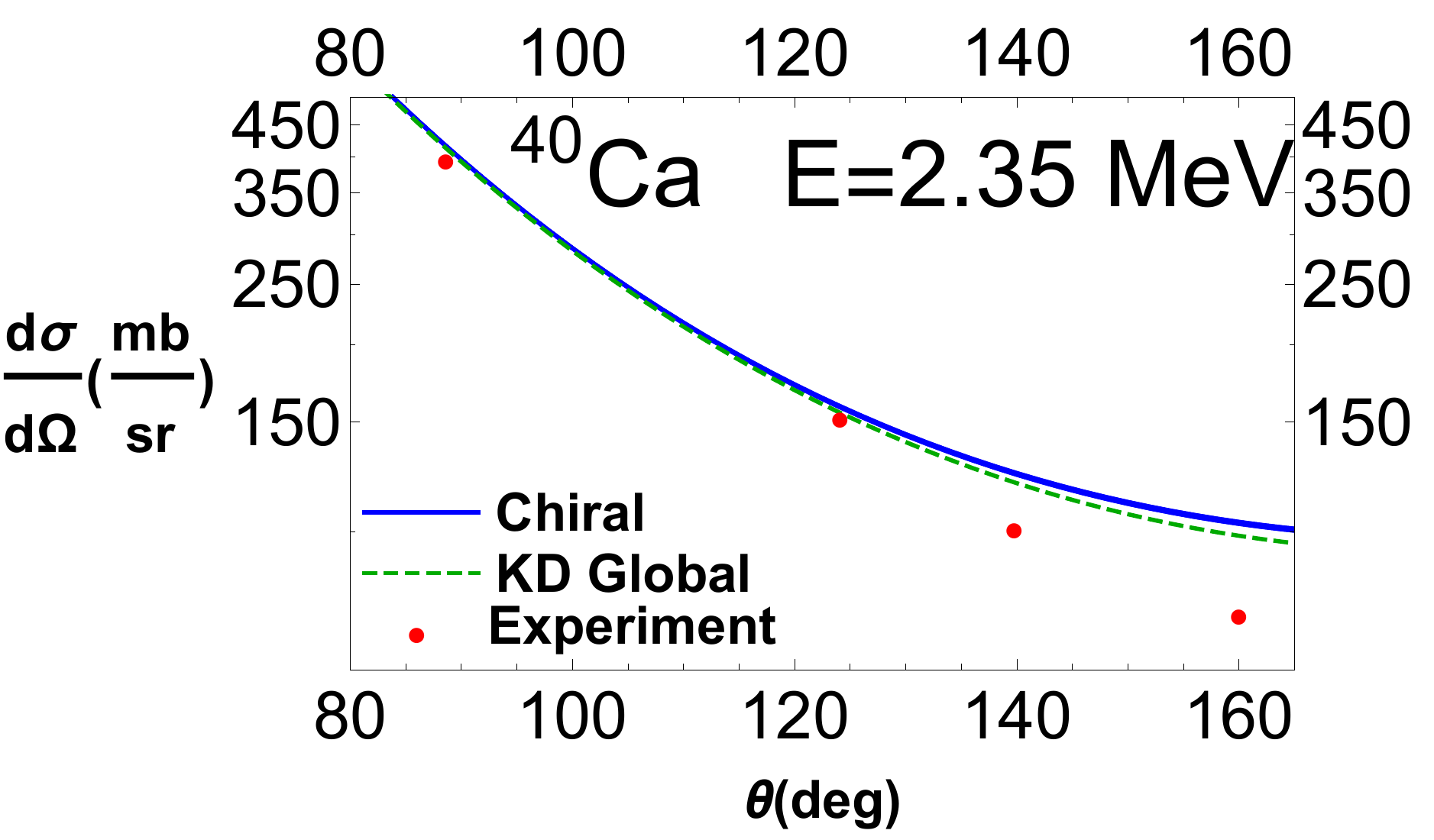}
\caption{Differential elastic scattering cross sections for proton-$^{40}$Ca at projectile energy $E=2.35$\,MeV. The microscopic cross section is given by the blue band. The KD phenomenological cross section is given by the green dashed curve, and experimental data are represented by red circles.}
\label{csplot}
\end{figure}

\section{Results}
\label{results}

\begin{flushleft}
\begin{figure*}[t]
\includegraphics[scale=0.299]{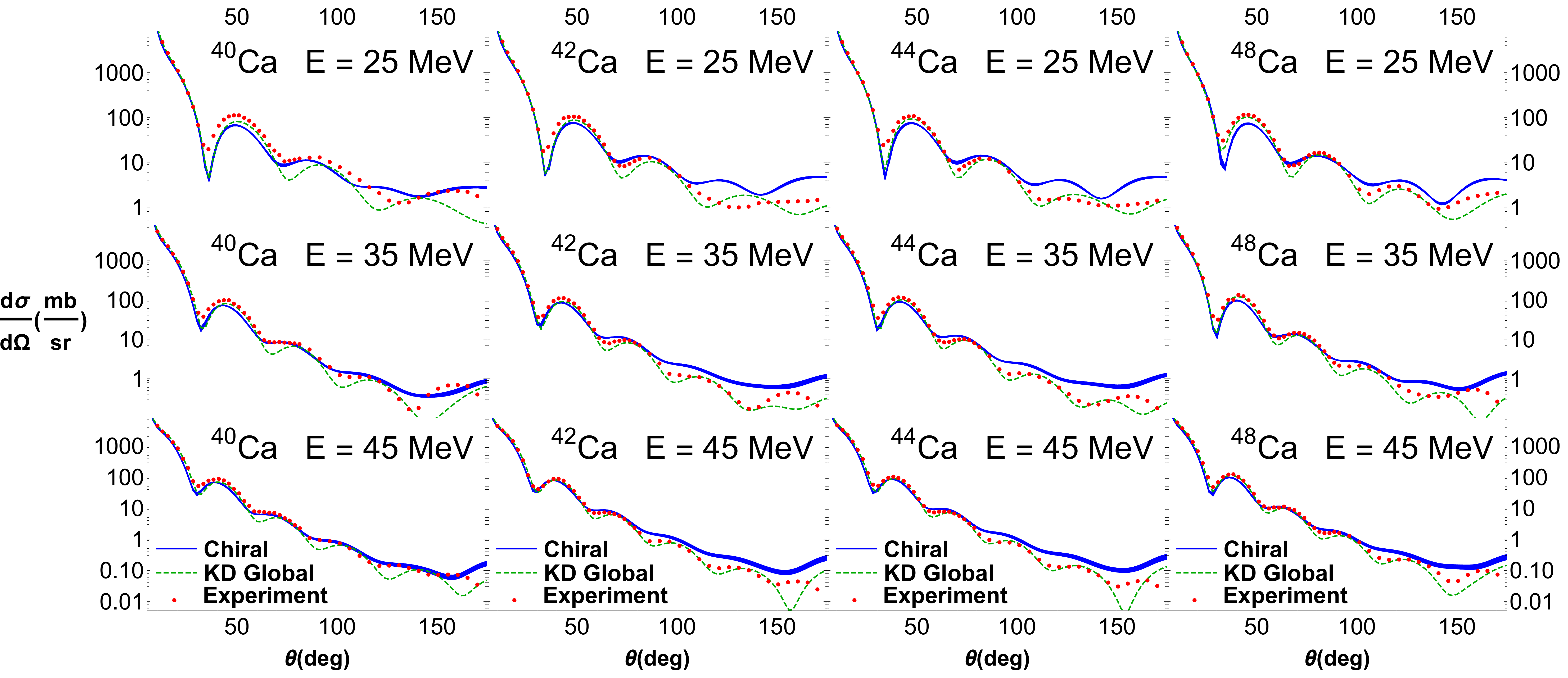}
\caption{Differential elastic scattering cross sections for proton projectiles on calcium targets at the energies $E=25,35,45$\,MeV. The microscopic cross sections are shown as the blue band. The KD phenomenological cross sections are given by the green dashed curves and experimental data are represented by red circles.}
\label{csplot1}
\end{figure*}

\begin{figure*}[t] 
\includegraphics[scale=0.299]{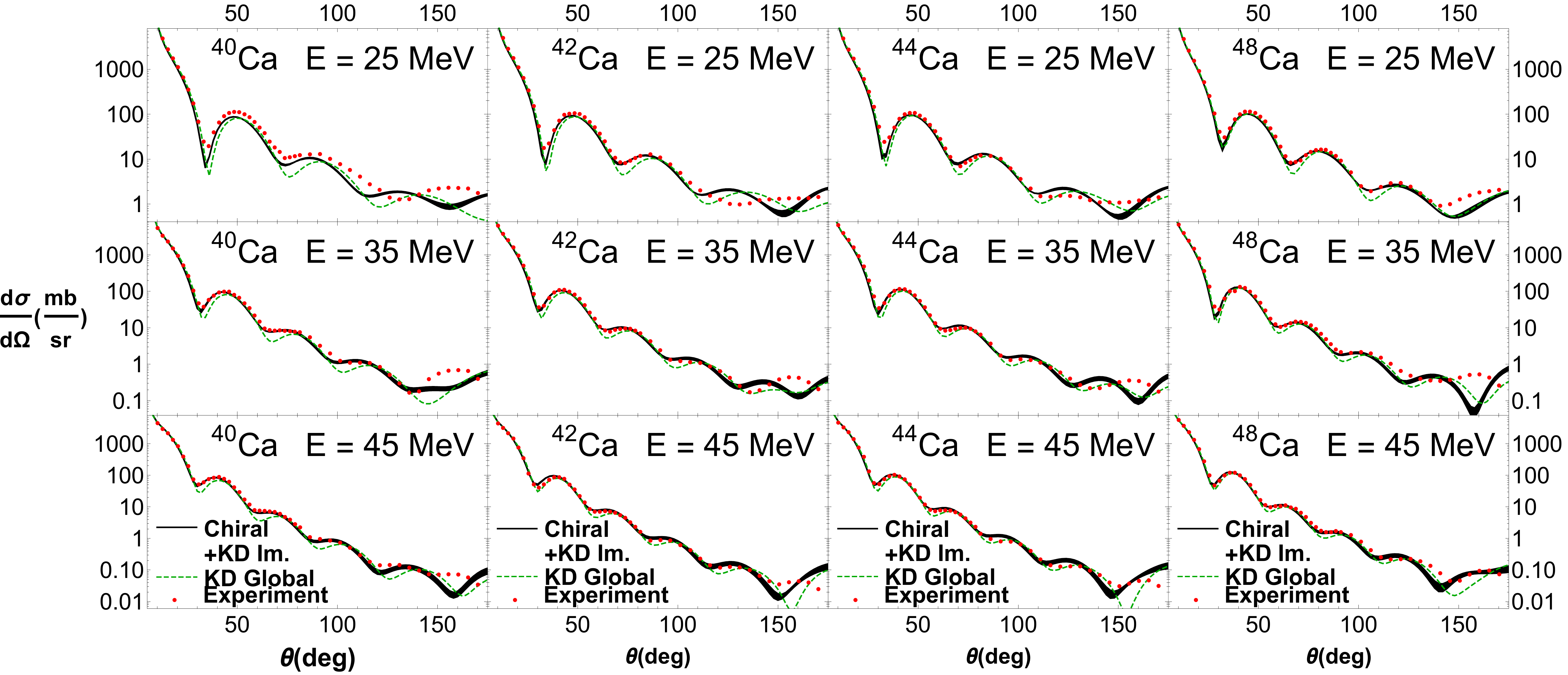}
\caption{Differential elastic scattering cross sections for proton projectiles on calcium targets at the energies $E=25,35,45$\,MeV. In comparison to Fig.\ \ref{csplot1}, the black bands show the cross sections that result from replacing the microscopic imaginary part in the chiral optical potential by the Koning-Delaroche phenomenological imaginary part.}
\label{csplot2}
\end{figure*}
\end{flushleft}

As a first test of the microscopic optical potentials constructed in the present work, we consider proton scattering on calcium isotopes. Both the differential elastic scattering cross sections and total reaction cross sections are calculated for selected calcium isotopes at energies for which there are available experimental data. In particular, we compute differential elastic scattering cross sections for $^{40,42,44,48}$Ca targets at $E=25, 35, 45 \: \text{MeV}$ projectile energies. For $^{40}$Ca, differential elastic scattering cross sections are also calculated at $E=2.35, 55, 65, 80, 135, 160 \: \text{MeV}$. Experimental data are taken from Refs.\ \cite{Calciumdata,Ca40e55,Ca4065,Ca40e80160,KOLTAY1975173}. The total reaction cross sections for proton scattering on $^{40,42,44,48}$Ca are calculated and compared to experimental data \cite{reactionCSCa40,reactionCSCa40low,reactionCSCa40mid,reactionCSCa424448}. Energies exceeding 200 MeV are not considered, since the chiral expansion is expected to breakdown around that energy scale.

The TALYS reaction code is used to calculate the cross sections in the different reaction channels. In all cases we employ the microscopic optical potential parametrized to the KD form implemented in the ILDA. In the present work, the only theoretical uncertainties considered are those for the ILDA length scales $t_C$ and $t_{SO}$. In the future, we will consider a wider class of chiral nuclear potentials in order to more accurately assess the complete theoretical uncertainty. We also benchmark against results from the KD global phenomenological optical potential \cite{KD03}. 

\subsection{Microscopic optical potential at low energy}
Low-energy nuclear reactions are important for a wide range of astrophysical applications. One of the primary motivations for the construction of new global microscopic optical potentials is to reduce the uncertainty in calculated radiative neutron capture reaction rates on exotic, neutron-rich isotopes. These reactions play an important role in $r$-process nucleosynthesis \cite{mumpower15,horowitz18}, especially in cold $r$-process environments such as neutron star mergers where freeze-out is achieved rapidly and neutron capture plays an enhanced role. Neutron-capture rates are included in most modern $r$-process reaction network codes, and the neutron-nucleus optical potential (together with level densities and $\gamma$ transition strength functions) is one of the key ingredients for the theoretical calculations. Most relevant is the imaginary part of the optical potential at low energies \cite{goriely07}.

In Fig.\ \ref{csplot} we show the differential elastic scattering cross sections for proton projectiles on a $^{40}$Ca target at $E=2.35$ MeV. The red circles indicate experimental data \cite{Calciumdata,Ca40e55,Ca4065,Ca40e80160,KOLTAY1975173}, the green dashed curve is the result of the global phenomenological optical potential from Koning and Delaroche, while the blue band is the prediction from the microscopic optical potential constructed in the present work.  Interestingly, there is very little difference between the phenomenological optical potential predictions and those from chiral effective field theory. Both calculations agree well with experimental data at scattering angles up to $\theta \simeq 120^\circ$, but overpredict the cross section at large angles.

\subsection{Microscopic optical potential at medium energy}
In Fig.\ \ref{csplot1} we plot the differential elastic scattering cross sections for protons on $^{40,42,44,48}$Ca targets at $E = 24, 35, 45$\,MeV. For scattering angles in the range $0^\circ$ $ < \theta < $ 80$^\circ$, the microscopic optical potential yields cross sections that are consistent with experiment and often more accurate than predictions based on the phenomenological KD optical potential. However, at larger scattering angles the microscopic calculations of the cross sections exhibit a weaker interference pattern, which persists as the energy increases. Overall, the microscopic elastic scattering cross sections are larger than experiment at high scattering angles.

\begin{figure}[t] 
\includegraphics[scale=0.34]{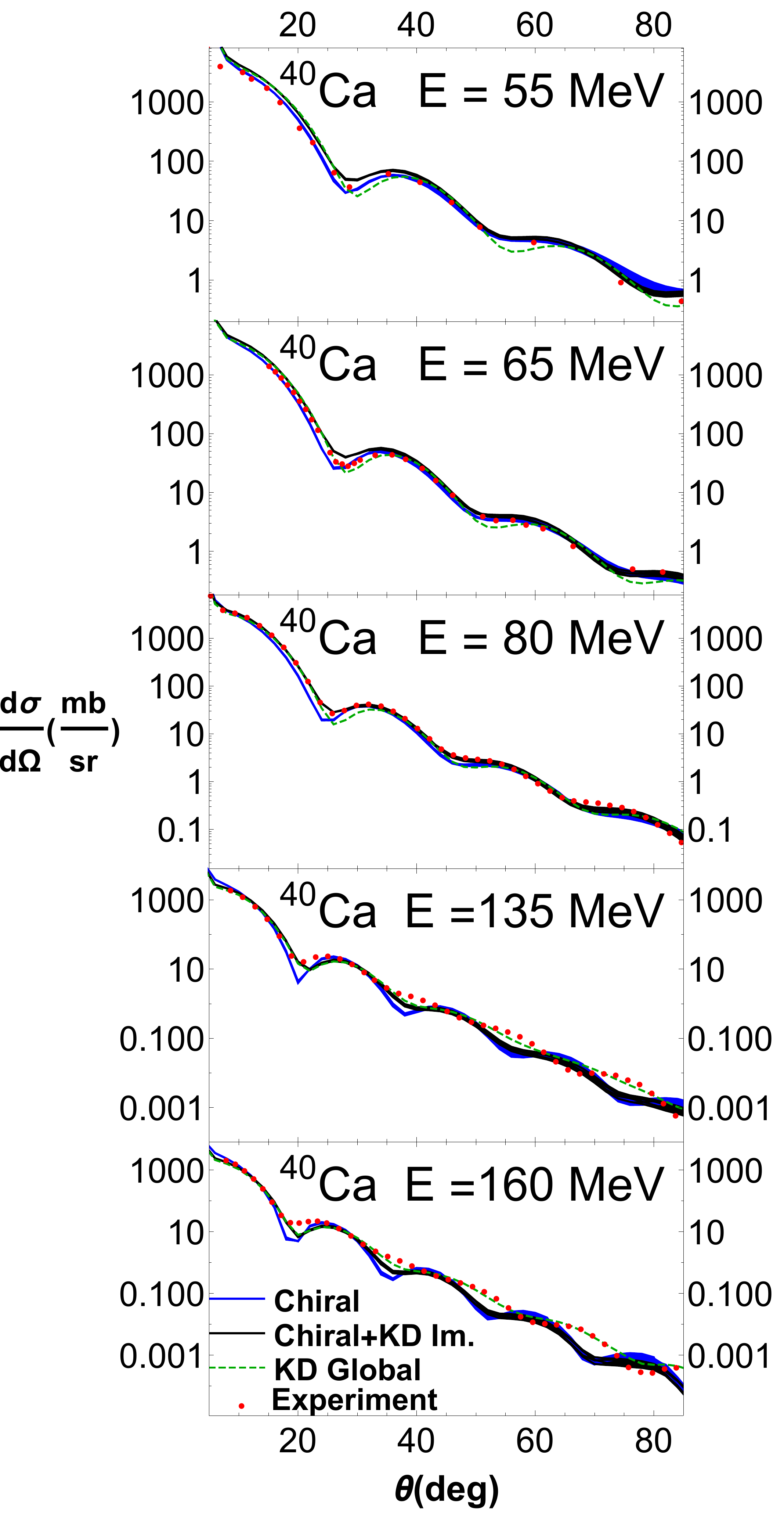}
\caption{Differential elastic scattering cross sections for proton projectiles on a $^{40}$Ca target at the energies $E=55,65,80,135,160$\,MeV. Full microscopic cross sections are shown as the blue bands, microscopic real optical potential plus phenomenological imaginary optical potential are shown by the black bands, the KD phenomenological cross sections are given by the green dashed curves, and experimental data are represented by red circles.}
\label{csplot3}
\end{figure} 

From Fig.\ \ref{fig:potentialplots35}, we suspect that the underlying cause of these discrepancies may be due to the imaginary part of the microscopic optical potential. At these intermediate projectile energies, the imaginary volume integral is close to phenomenology according to Table \ref{table:VolumeIntegrals}. However, the microscopic surface imaginary peak is too small, as can be seen in Fig.\ \ref{fig:potentialplots35}, which leads to larger elastic scattering cross sections. In contrast the imaginary volume part, probed at higher projectile energies, is much larger than phenomenology.

In order to investigate this conjecture, we substitute the phenomenological imaginary term into the microscopic optical potential. This replacement is meant to be a simple way of showing the possible improvements in the chiral optical potential and should not be interpreted as a substitute for proper microscopic modeling. In Fig.\ \ref{csplot2} we show the differential elastic scattering cross sections for protons on $^{40,42,44,48}$Ca targets at $E = 24, 35, 45$\,MeV with this phenomenological replacement. Indeed we find that the calculated cross sections are much improved at large angles across all isotopes. The enhanced surface imaginary part leads to stronger interferences and an overall decrease in the elastic scattering cross section. Hence, there is a strong motivation for future work aimed at improving the microscopic description of the imaginary part of the optical potential.

\subsection{Microscopic optical potential at high energy}
To test the chiral optical potential at higher energies, we calculate proton-$^{40}$Ca differential elastic scattering cross sections at $E=55, 65, 80, 135, 160$\,MeV. In Fig.\ \ref{csplot3} we plot the results from the chiral optical potential and the KD phenomenological optical potential together with experimental data from Refs.\ \cite{Calciumdata,Ca40e55,Ca4065,Ca40e80160,KOLTAY1975173}. The cross sections from the chiral optical potential stay close to phenomenological and experimental results for $E = 55, 65$ MeV but begin to deviate strongly for $E > 80$\,MeV. The microscopic imaginary term becomes much more absorptive for $E > 80$\,MeV, as the large volume contribution from the chiral optical potential becomes more relevant. The effect of this can be seen especially in the lower three plots of Fig.\ \ref{csplot3}, where the cross sections exhibit large interference oscillations. Since there are more open inelastic channels at higher energy, a stronger imaginary part in general corresponds to a lower elastic scattering cross section.

\begin{figure}[t] 
\includegraphics[scale=0.48]{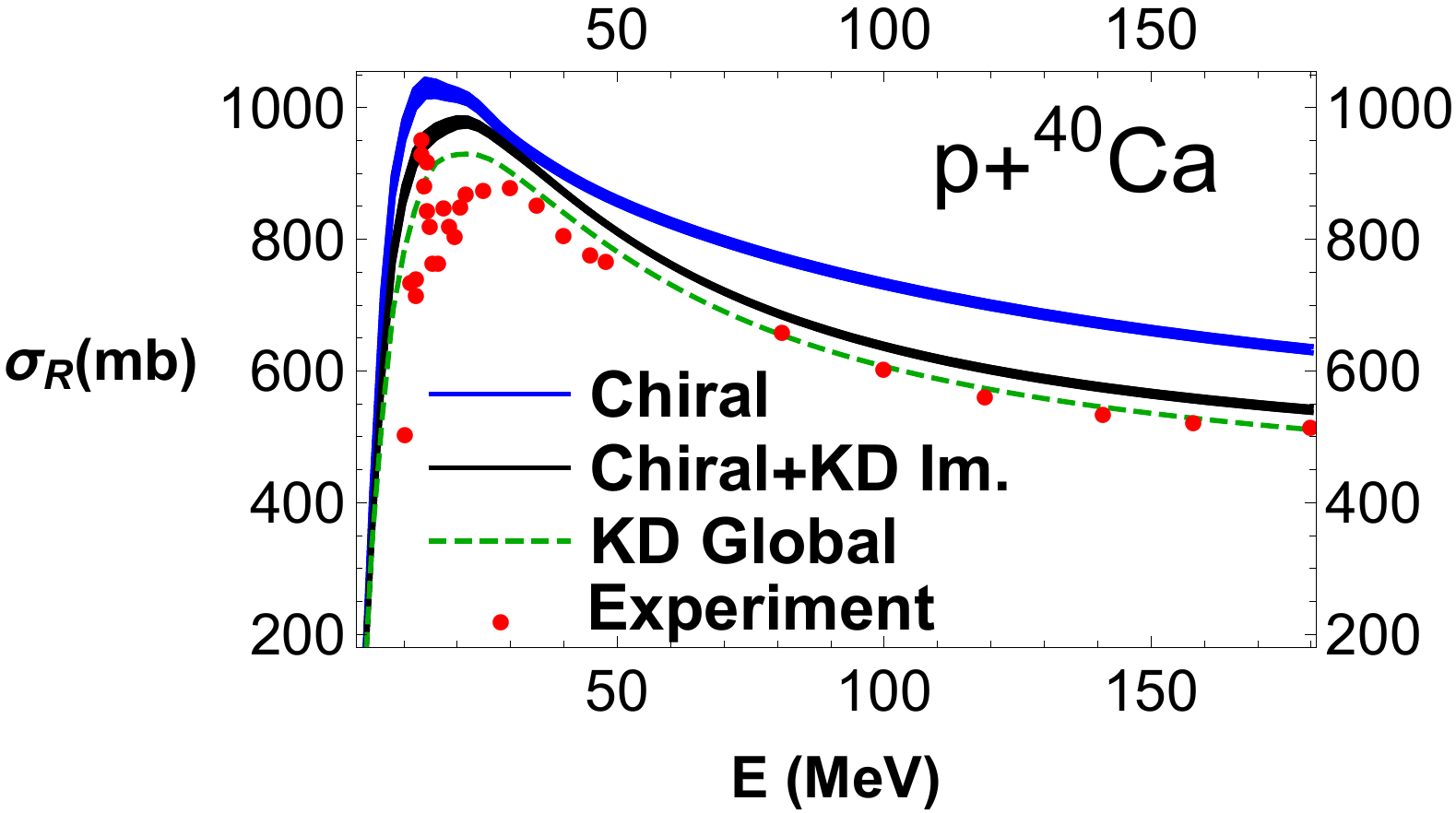}
\caption{The proton-$^{40}$Ca total reaction cross section as a function of energy calculated from the microscopic optical potential (blue band), the phenomenological KD optical potential (dashed green curve), and the microscopic optical potential with the phenomenological KD imaginary part (black band). Experimental data are shown as red circles.}
\label{totalcsfig40}
\end{figure} 

\begin{figure}[t!] 
\includegraphics[scale=0.44]{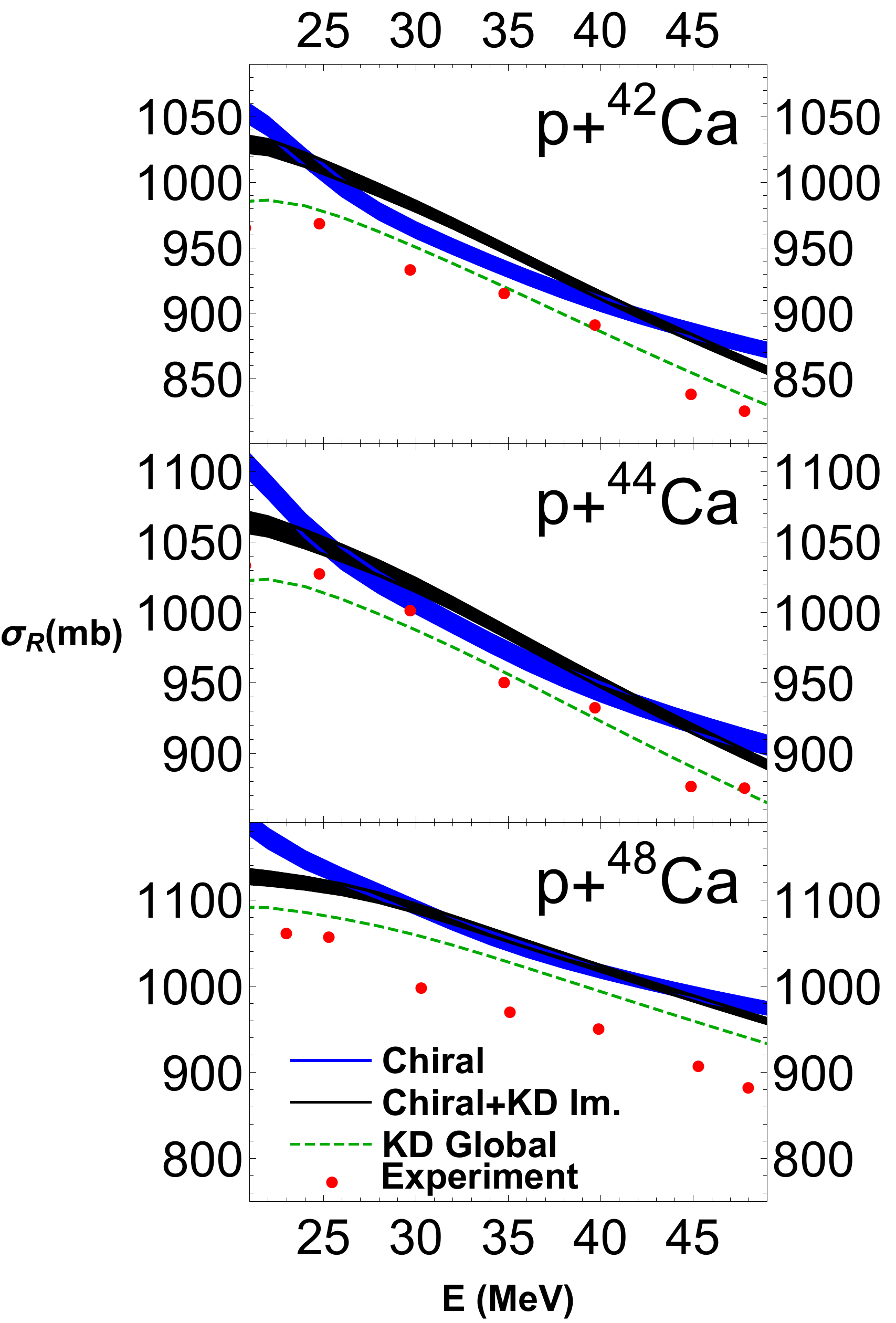}
\caption{The proton-$^{42,44,48}$Ca total reaction cross sections from $20\,{\rm MeV} < E < 50\,{\rm MeV}$ calculated from the microscopic optical potential (blue bands), the phenomenological KD optical potential (dashed green curves), and the microscopic optical potential with the phenomenological KD imaginary part (black bands). Experimental data are shown as red circles.}
\label{totalcsfig}
\end{figure} 

In order to assess the quality of the microscopic imaginary part of the optical potential, we again substitute the KD phenomenological imaginary part into the chiral optical potential. The results for proton-$^{40}$Ca elastic scattering cross sections at $E=55, 65, 80, 135, 160$\,MeV are shown in Fig.\ \ref{csplot3}. Again, we find that the replacement of the large microscopic imaginary optical potential by the KD phenomenological imaginary part leads to significant improvements in the elastic scattering cross sections across all energies. For $E = 135$ and $160$ MeV, the purely phenomenological cross sections are still more accurate, but the microscopic optical potential with phenomenological imaginary part gives a quality description of the data.

\subsection{Total reaction cross section}
In Figs.\ \ref{totalcsfig40} and \ref{totalcsfig} we plot the total reaction cross sections for proton scattering on $^{40,42,44,48}$Ca from our microscopic optical potential and the KD phenomenological optical potential. The chiral EFT results are shown as the blue band, while the KD predictions are shown as dashed green lines. Experimental data \cite{reactionCSCa40,reactionCSCa40low,reactionCSCa40mid,reactionCSCa424448} are shown with red circles. We see that for all energies the purely microscopic optical potential predicts a total reaction cross section that is too large, while the phenomenological potential gives an overall good description for most isotopes and energies. However, for proton-$^{48}$Ca, the KD phenomenological potential also gives a larger total reaction cross section compared to experiment. This suggests that there might be room for improvement in the phenomenological description of the global isovector optical potential probed in this neutron-rich nucleus.

We also show in Figs.\ \ref{totalcsfig40} and \ref{totalcsfig} the results for the total reaction cross sections (black solid bands) when the microscopic imaginary part is replaced by the KD phenomenological imaginary potential. For intermediate energies ($20\,{\rm MeV} < E < 50\,{\rm MeV}$), there is not a substantial improvement in the comparison to experimental data. However, beyond energies of $E=50$\,MeV that are shown in Fig.\ \ref{totalcsfig40}, the replacement of the phenomenological imaginary part again leads to a significant improvement in the description of the total reaction cross section. Nevertheless, the modified microscopic optical potential still overestimates the total reaction cross section for all energies due to the real volume and real spin-orbit terms having slightly larger depths than their phenomenological counterparts.

\section{Conclusions}
\label{conclusions}
We have calculated a microscopic optical potential from chiral two- and three-body forces for proton scattering on calcium isotopes. We started from a self-consistent second-order calculation of the proton and neutron self energies in isospin-asymmetric nuclear matter from which we derived the central real and imaginary parts of the optical potential in finite nuclei within the framework of the improved local density approximation. The real spin-orbit potential was constructed from the improved density matrix expansion using the same chiral two- and three-body forces.

We found that chiral nucleon-nucleus optical potentials describe low-energy ($E \lesssim 5$\,MeV) scattering processes rather well, due in part to a well-defined surface peak in the imaginary part of the optical potential. At all energies, the real central term is consistent with phenomenological modeling, while the microscopic spin-orbit strength is larger by $\sim 20$\%.  At moderate energies ($E \gtrsim 35$\,MeV), the imaginary part of the chiral optical potential develops a large volume term without a significant surface peak. This leads to discrepancies between our theoretical calculations and experimental data, especially at large scattering angles. At the highest energies ($E \simeq100-160$\,MeV) considered in the present work, the large imaginary term leads to over suppression of the elastic scattering cross section. We have shown that substituting the microscopic imaginary part with the KD phenomenological optical potential leads to excellent agreement with elastic scattering and total reaction cross sections for nearly all isotopes and projectile energies investigated. In the future we plan to compute higher-order perturbative contributions that may improve the description of the imaginary part of the optical potential and the overall spin-orbit strength. We will also explore a wider range \cite{Sammarruca18} of chiral nuclear potentials in order to provide a more comprehensive estimate of the theoretical uncertainties.

\vspace{.5in}
\begin{acknowledgments}
We thank Carlos Bertulani for useful discussions and L. Huff for comments on the manuscript. Work supported by the National Science Foundation under Grant No. PHY1652199 and by the U.S. Department of Energy National Nuclear Security Administration under Grant No. DE-NA0003841. Portions of this research were conducted with the advanced computing resources provided by Texas A\&M High Performance Research Computing.
\end{acknowledgments}

\bibliographystyle{apsrev4-1}

\begin{thebibliography}{76}%
\makeatletter
\providecommand \@ifxundefined [1]{%
 \@ifx{#1\undefined}
}%
\providecommand \@ifnum [1]{%
 \ifnum #1\expandafter \@firstoftwo
 \else \expandafter \@secondoftwo
 \fi
}%
\providecommand \@ifx [1]{%
 \ifx #1\expandafter \@firstoftwo
 \else \expandafter \@secondoftwo
 \fi
}%
\providecommand \natexlab [1]{#1}%
\providecommand \enquote  [1]{``#1''}%
\providecommand \bibnamefont  [1]{#1}%
\providecommand \bibfnamefont [1]{#1}%
\providecommand \citenamefont [1]{#1}%
\providecommand \href@noop [0]{\@secondoftwo}%
\providecommand \href [0]{\begingroup \@sanitize@url \@href}%
\providecommand \@href[1]{\@@startlink{#1}\@@href}%
\providecommand \@@href[1]{\endgroup#1\@@endlink}%
\providecommand \@sanitize@url [0]{\catcode `\\12\catcode `\$12\catcode
  `\&12\catcode `\#12\catcode `\^12\catcode `\_12\catcode `\%12\relax}%
\providecommand \@@startlink[1]{}%
\providecommand \@@endlink[0]{}%
\providecommand \url  [0]{\begingroup\@sanitize@url \@url }%
\providecommand \@url [1]{\endgroup\@href {#1}{\urlprefix }}%
\providecommand \urlprefix  [0]{URL }%
\providecommand \Eprint [0]{\href }%
\providecommand \doibase [0]{http://dx.doi.org/}%
\providecommand \selectlanguage [0]{\@gobble}%
\providecommand \bibinfo  [0]{\@secondoftwo}%
\providecommand \bibfield  [0]{\@secondoftwo}%
\providecommand \translation [1]{[#1]}%
\providecommand \BibitemOpen [0]{}%
\providecommand \bibitemStop [0]{}%
\providecommand \bibitemNoStop [0]{.\EOS\space}%
\providecommand \EOS [0]{\spacefactor3000\relax}%
\providecommand \BibitemShut  [1]{\csname bibitem#1\endcsname}%
\let\auto@bib@innerbib\@empty
\bibitem [{\citenamefont {Varner}\ \emph {et~al.}(1991)\citenamefont {Varner},
  \citenamefont {Thompson}, \citenamefont {McAbee}, \citenamefont {Ludwig},\
  and\ \citenamefont {Clegg}}]{Varner91}%
  \BibitemOpen
  \bibfield  {author} {\bibinfo {author} {\bibfnamefont {R.~L.}\ \bibnamefont
  {Varner}}, \bibinfo {author} {\bibfnamefont {W.~J.}\ \bibnamefont
  {Thompson}}, \bibinfo {author} {\bibfnamefont {T.~L.}\ \bibnamefont
  {McAbee}}, \bibinfo {author} {\bibfnamefont {E.~J.}\ \bibnamefont {Ludwig}},
  \ and\ \bibinfo {author} {\bibfnamefont {T.~B.}\ \bibnamefont {Clegg}},\
  }\href {\doibase https://doi.org/10.1016/0370-1573(91)90039-O} {\bibfield
  {journal} {\bibinfo  {journal} {Phys. Rep.}\ }\textbf {\bibinfo {volume}
  {201}},\ \bibinfo {pages} {57 } (\bibinfo {year} {1991})}\BibitemShut
  {NoStop}%
\bibitem [{\citenamefont {Koning}\ and\ \citenamefont
  {Delaroche}(2003)}]{KD03}%
  \BibitemOpen
  \bibfield  {author} {\bibinfo {author} {\bibfnamefont {A.~J.}\ \bibnamefont
  {Koning}}\ and\ \bibinfo {author} {\bibfnamefont {J.~P.}\ \bibnamefont
  {Delaroche}},\ }\href {\doibase
  https://doi.org/10.1016/S0375-9474(02)01321-0} {\bibfield  {journal}
  {\bibinfo  {journal} {Nucl. Phys.}\ }\textbf {\bibinfo {volume} {A713}},\
  \bibinfo {pages} {231 } (\bibinfo {year} {2003})}\BibitemShut {NoStop}%
\bibitem [{\citenamefont {Jeukenne}\ \emph {et~al.}(1976)\citenamefont
  {Jeukenne}, \citenamefont {Lejeune},\ and\ \citenamefont
  {Mahaux}}]{JEUKENNE1976}%
  \BibitemOpen
  \bibfield  {author} {\bibinfo {author} {\bibfnamefont {J.~P.}\ \bibnamefont
  {Jeukenne}}, \bibinfo {author} {\bibfnamefont {A.}~\bibnamefont {Lejeune}}, \
  and\ \bibinfo {author} {\bibfnamefont {C.}~\bibnamefont {Mahaux}},\ }\href
  {\doibase https://doi.org/10.1016/0370-1573(76)90017-X} {\bibfield  {journal}
  {\bibinfo  {journal} {Phys. Rep.}\ }\textbf {\bibinfo {volume} {25}},\
  \bibinfo {pages} {83 } (\bibinfo {year} {1976})}\BibitemShut {NoStop}%
\bibitem [{\citenamefont {Lejeune}\ and\ \citenamefont
  {Hodgson}(1978)}]{LEJEUNE78}%
  \BibitemOpen
  \bibfield  {author} {\bibinfo {author} {\bibfnamefont {A.}~\bibnamefont
  {Lejeune}}\ and\ \bibinfo {author} {\bibfnamefont {P.~E.}\ \bibnamefont
  {Hodgson}},\ }\href {\doibase https://doi.org/10.1016/0375-9474(78)90118-5}
  {\bibfield  {journal} {\bibinfo  {journal} {Nucl. Phys.}\ }\textbf {\bibinfo
  {volume} {A295}},\ \bibinfo {pages} {301 } (\bibinfo {year}
  {1978})}\BibitemShut {NoStop}%
\bibitem [{\citenamefont {Hansen}\ \emph {et~al.}(1985)\citenamefont {Hansen},
  \citenamefont {Dietrich}, \citenamefont {Pohl}, \citenamefont {Poppe},\ and\
  \citenamefont {Wong}}]{Hansen85}%
  \BibitemOpen
  \bibfield  {author} {\bibinfo {author} {\bibfnamefont {L.~F.}\ \bibnamefont
  {Hansen}}, \bibinfo {author} {\bibfnamefont {F.~S.}\ \bibnamefont
  {Dietrich}}, \bibinfo {author} {\bibfnamefont {B.~A.}\ \bibnamefont {Pohl}},
  \bibinfo {author} {\bibfnamefont {C.~H.}\ \bibnamefont {Poppe}}, \ and\
  \bibinfo {author} {\bibfnamefont {C.}~\bibnamefont {Wong}},\ }\href {\doibase
  10.1103/PhysRevC.31.111} {\bibfield  {journal} {\bibinfo  {journal} {Phys.
  Rev. C}\ }\textbf {\bibinfo {volume} {31}},\ \bibinfo {pages} {111} (\bibinfo
  {year} {1985})}\BibitemShut {NoStop}%
\bibitem [{\citenamefont {Petler}\ \emph {et~al.}(1985)\citenamefont {Petler},
  \citenamefont {Islam}, \citenamefont {Finlay},\ and\ \citenamefont
  {Dietrich}}]{Petler85}%
  \BibitemOpen
  \bibfield  {author} {\bibinfo {author} {\bibfnamefont {J.~S.}\ \bibnamefont
  {Petler}}, \bibinfo {author} {\bibfnamefont {M.~S.}\ \bibnamefont {Islam}},
  \bibinfo {author} {\bibfnamefont {R.~W.}\ \bibnamefont {Finlay}}, \ and\
  \bibinfo {author} {\bibfnamefont {F.~S.}\ \bibnamefont {Dietrich}},\ }\href
  {\doibase 10.1103/PhysRevC.32.673} {\bibfield  {journal} {\bibinfo  {journal}
  {Phys. Rev. C}\ }\textbf {\bibinfo {volume} {32}},\ \bibinfo {pages} {673}
  (\bibinfo {year} {1985})}\BibitemShut {NoStop}%
\bibitem [{\citenamefont {Bauge}\ \emph {et~al.}(1998)\citenamefont {Bauge},
  \citenamefont {Delaroche},\ and\ \citenamefont {Girod}}]{DelarocheILDA}%
  \BibitemOpen
  \bibfield  {author} {\bibinfo {author} {\bibfnamefont {E.}~\bibnamefont
  {Bauge}}, \bibinfo {author} {\bibfnamefont {J.~P.}\ \bibnamefont
  {Delaroche}}, \ and\ \bibinfo {author} {\bibfnamefont {M.}~\bibnamefont
  {Girod}},\ }\href {\doibase 10.1103/PhysRevC.58.1118} {\bibfield  {journal}
  {\bibinfo  {journal} {Phys. Rev. C}\ }\textbf {\bibinfo {volume} {58}},\
  \bibinfo {pages} {1118} (\bibinfo {year} {1998})}\BibitemShut {NoStop}%
\bibitem [{\citenamefont {Bauge}\ \emph {et~al.}(2001)\citenamefont {Bauge},
  \citenamefont {Delaroche},\ and\ \citenamefont {Girod}}]{bauge01}%
  \BibitemOpen
  \bibfield  {author} {\bibinfo {author} {\bibfnamefont {E.}~\bibnamefont
  {Bauge}}, \bibinfo {author} {\bibfnamefont {J.~P.}\ \bibnamefont
  {Delaroche}}, \ and\ \bibinfo {author} {\bibfnamefont {M.}~\bibnamefont
  {Girod}},\ }\href@noop {} {\bibfield  {journal} {\bibinfo  {journal} {Phys.
  Rev. C}\ }\textbf {\bibinfo {volume} {63}},\ \bibinfo {pages} {024607}
  (\bibinfo {year} {2001})}\BibitemShut {NoStop}%
\bibitem [{\citenamefont {Goriely}\ and\ \citenamefont
  {Delaroche}(2007)}]{goriely07}%
  \BibitemOpen
  \bibfield  {author} {\bibinfo {author} {\bibfnamefont {S.}~\bibnamefont
  {Goriely}}\ and\ \bibinfo {author} {\bibfnamefont {J.-P.}\ \bibnamefont
  {Delaroche}},\ }\href@noop {} {\bibfield  {journal} {\bibinfo  {journal}
  {Phys. Lett.}\ }\textbf {\bibinfo {volume} {B653}},\ \bibinfo {pages} {178}
  (\bibinfo {year} {2007})}\BibitemShut {NoStop}%
\bibitem [{\citenamefont {Holt}\ \emph {et~al.}(2013)\citenamefont {Holt},
  \citenamefont {Kaiser}, \citenamefont {Miller},\ and\ \citenamefont
  {Weise}}]{Holt13omp}%
  \BibitemOpen
  \bibfield  {author} {\bibinfo {author} {\bibfnamefont {J.~W.}\ \bibnamefont
  {Holt}}, \bibinfo {author} {\bibfnamefont {N.}~\bibnamefont {Kaiser}},
  \bibinfo {author} {\bibfnamefont {G.~A.}\ \bibnamefont {Miller}}, \ and\
  \bibinfo {author} {\bibfnamefont {W.}~\bibnamefont {Weise}},\ }\href
  {\doibase 10.1103/PhysRevC.88.024614} {\bibfield  {journal} {\bibinfo
  {journal} {Phys. Rev. C}\ }\textbf {\bibinfo {volume} {88}},\ \bibinfo
  {pages} {024614} (\bibinfo {year} {2013})}\BibitemShut {NoStop}%
\bibitem [{\citenamefont {Toyokawa}\ \emph
  {et~al.}(2015{\natexlab{a}})\citenamefont {Toyokawa}, \citenamefont {Minomo},
  \citenamefont {Kohno},\ and\ \citenamefont {Yahiro}}]{Toyokawa3N}%
  \BibitemOpen
  \bibfield  {author} {\bibinfo {author} {\bibfnamefont {M.}~\bibnamefont
  {Toyokawa}}, \bibinfo {author} {\bibfnamefont {K.}~\bibnamefont {Minomo}},
  \bibinfo {author} {\bibfnamefont {M.}~\bibnamefont {Kohno}}, \ and\ \bibinfo
  {author} {\bibfnamefont {M.}~\bibnamefont {Yahiro}},\ }\href
  {http://stacks.iop.org/0954-3899/42/i=2/a=025104} {\bibfield  {journal}
  {\bibinfo  {journal} {J. Phys. G}\ }\textbf {\bibinfo {volume} {42}},\
  \bibinfo {pages} {025104} (\bibinfo {year} {2015}{\natexlab{a}})}\BibitemShut
  {NoStop}%
\bibitem [{\citenamefont {Holt}\ \emph
  {et~al.}(2016{\natexlab{a}})\citenamefont {Holt}, \citenamefont {Kaiser},\
  and\ \citenamefont {Miller}}]{Holt15omp}%
  \BibitemOpen
  \bibfield  {author} {\bibinfo {author} {\bibfnamefont {J.~W.}\ \bibnamefont
  {Holt}}, \bibinfo {author} {\bibfnamefont {N.}~\bibnamefont {Kaiser}}, \ and\
  \bibinfo {author} {\bibfnamefont {G.~A.}\ \bibnamefont {Miller}},\ }\href
  {\doibase 10.1103/PhysRevC.93.064603} {\bibfield  {journal} {\bibinfo
  {journal} {Phys. Rev. C}\ }\textbf {\bibinfo {volume} {93}},\ \bibinfo
  {pages} {064603} (\bibinfo {year} {2016}{\natexlab{a}})}\BibitemShut
  {NoStop}%
\bibitem [{\citenamefont {Camarda}\ \emph {et~al.}(1989)\citenamefont
  {Camarda}, \citenamefont {Dietrich},\ and\ \citenamefont
  {Phillips}}]{camarda89}%
  \BibitemOpen
  \bibfield  {author} {\bibinfo {author} {\bibfnamefont {H.~S.}\ \bibnamefont
  {Camarda}}, \bibinfo {author} {\bibfnamefont {F.~S.}\ \bibnamefont
  {Dietrich}}, \ and\ \bibinfo {author} {\bibfnamefont {T.~W.}\ \bibnamefont
  {Phillips}},\ }\href@noop {} {\bibfield  {journal} {\bibinfo  {journal}
  {Phys. Rev. C}\ }\textbf {\bibinfo {volume} {39}},\ \bibinfo {pages} {1725}
  (\bibinfo {year} {1989})}\BibitemShut {NoStop}%
\bibitem [{\citenamefont {Kerman}\ \emph {et~al.}(1959)\citenamefont {Kerman},
  \citenamefont {McManus},\ and\ \citenamefont {Thaler}}]{kerman59}%
  \BibitemOpen
  \bibfield  {author} {\bibinfo {author} {\bibfnamefont {A.~K.}\ \bibnamefont
  {Kerman}}, \bibinfo {author} {\bibfnamefont {H.}~\bibnamefont {McManus}}, \
  and\ \bibinfo {author} {\bibfnamefont {R.}~\bibnamefont {Thaler}},\
  }\href@noop {} {\bibfield  {journal} {\bibinfo  {journal} {Ann. Phys.
  (N.Y.)}\ }\textbf {\bibinfo {volume} {8}},\ \bibinfo {pages} {551} (\bibinfo
  {year} {1959})}\BibitemShut {NoStop}%
\bibitem [{\citenamefont {Brieva}\ and\ \citenamefont
  {Rook}(1977{\natexlab{a}})}]{BRIEVA1977299}%
  \BibitemOpen
  \bibfield  {author} {\bibinfo {author} {\bibfnamefont {F.~A.}\ \bibnamefont
  {Brieva}}\ and\ \bibinfo {author} {\bibfnamefont {J.~R.}\ \bibnamefont
  {Rook}},\ }\href {\doibase https://doi.org/10.1016/0375-9474(77)90322-0}
  {\bibfield  {journal} {\bibinfo  {journal} {Nucl. Phys.}\ }\textbf {\bibinfo
  {volume} {A291}},\ \bibinfo {pages} {299 } (\bibinfo {year}
  {1977}{\natexlab{a}})}\BibitemShut {NoStop}%
\bibitem [{\citenamefont {Ray}\ and\ \citenamefont
  {Hoffmann}(1985)}]{Hoffmann85}%
  \BibitemOpen
  \bibfield  {author} {\bibinfo {author} {\bibfnamefont {L.}~\bibnamefont
  {Ray}}\ and\ \bibinfo {author} {\bibfnamefont {G.~W.}\ \bibnamefont
  {Hoffmann}},\ }\href {\doibase 10.1103/PhysRevC.31.538} {\bibfield  {journal}
  {\bibinfo  {journal} {Phys. Rev. C}\ }\textbf {\bibinfo {volume} {31}},\
  \bibinfo {pages} {538} (\bibinfo {year} {1985})}\BibitemShut {NoStop}%
\bibitem [{\citenamefont {Elster}\ \emph {et~al.}(1990)\citenamefont {Elster},
  \citenamefont {Cheon}, \citenamefont {Redish},\ and\ \citenamefont
  {Tandy}}]{Elster90}%
  \BibitemOpen
  \bibfield  {author} {\bibinfo {author} {\bibfnamefont {C.}~\bibnamefont
  {Elster}}, \bibinfo {author} {\bibfnamefont {T.}~\bibnamefont {Cheon}},
  \bibinfo {author} {\bibfnamefont {E.~F.}\ \bibnamefont {Redish}}, \ and\
  \bibinfo {author} {\bibfnamefont {P.~C.}\ \bibnamefont {Tandy}},\ }\href
  {\doibase 10.1103/PhysRevC.41.814} {\bibfield  {journal} {\bibinfo  {journal}
  {Phys. Rev. C}\ }\textbf {\bibinfo {volume} {41}},\ \bibinfo {pages} {814}
  (\bibinfo {year} {1990})}\BibitemShut {NoStop}%
\bibitem [{\citenamefont {Arellano}\ \emph {et~al.}(1990)\citenamefont
  {Arellano}, \citenamefont {Brieva},\ and\ \citenamefont {Love}}]{Arellano90}%
  \BibitemOpen
  \bibfield  {author} {\bibinfo {author} {\bibfnamefont {H.~F.}\ \bibnamefont
  {Arellano}}, \bibinfo {author} {\bibfnamefont {F.~A.}\ \bibnamefont
  {Brieva}}, \ and\ \bibinfo {author} {\bibfnamefont {W.~G.}\ \bibnamefont
  {Love}},\ }\href {\doibase 10.1103/PhysRevC.41.2188} {\bibfield  {journal}
  {\bibinfo  {journal} {Phys. Rev. C}\ }\textbf {\bibinfo {volume} {41}},\
  \bibinfo {pages} {2188} (\bibinfo {year} {1990})}\BibitemShut {NoStop}%
\bibitem [{\citenamefont {Arellano}\ \emph {et~al.}(1995)\citenamefont
  {Arellano}, \citenamefont {Brieva},\ and\ \citenamefont {Love}}]{arellano95}%
  \BibitemOpen
  \bibfield  {author} {\bibinfo {author} {\bibfnamefont {H.~F.}\ \bibnamefont
  {Arellano}}, \bibinfo {author} {\bibfnamefont {F.~A.}\ \bibnamefont
  {Brieva}}, \ and\ \bibinfo {author} {\bibfnamefont {W.~G.}\ \bibnamefont
  {Love}},\ }\href@noop {} {\bibfield  {journal} {\bibinfo  {journal} {Phys.
  Rev. C}\ }\textbf {\bibinfo {volume} {52}},\ \bibinfo {pages} {301} (\bibinfo
  {year} {1995})}\BibitemShut {NoStop}%
\bibitem [{\citenamefont {Vorabbi}\ \emph {et~al.}(2016)\citenamefont
  {Vorabbi}, \citenamefont {Finelli},\ and\ \citenamefont
  {Giusti}}]{vorabbi16}%
  \BibitemOpen
  \bibfield  {author} {\bibinfo {author} {\bibfnamefont {M.}~\bibnamefont
  {Vorabbi}}, \bibinfo {author} {\bibfnamefont {P.}~\bibnamefont {Finelli}}, \
  and\ \bibinfo {author} {\bibfnamefont {C.}~\bibnamefont {Giusti}},\
  }\href@noop {} {\bibfield  {journal} {\bibinfo  {journal} {Phys. Rev. C}\
  }\textbf {\bibinfo {volume} {93}},\ \bibinfo {pages} {034619} (\bibinfo
  {year} {2016})}\BibitemShut {NoStop}%
\bibitem [{\citenamefont {Vorabbi}\ \emph {et~al.}(2018)\citenamefont
  {Vorabbi}, \citenamefont {Finelli},\ and\ \citenamefont
  {Giusti}}]{Vorabbi18}%
  \BibitemOpen
  \bibfield  {author} {\bibinfo {author} {\bibfnamefont {M.}~\bibnamefont
  {Vorabbi}}, \bibinfo {author} {\bibfnamefont {P.}~\bibnamefont {Finelli}}, \
  and\ \bibinfo {author} {\bibfnamefont {C.}~\bibnamefont {Giusti}},\
  }\href@noop {} {\bibfield  {journal} {\bibinfo  {journal} {Phys. Rev. C}\
  }\textbf {\bibinfo {volume} {98}},\ \bibinfo {pages} {064602} (\bibinfo
  {year} {2018})}\BibitemShut {NoStop}%
\bibitem [{\citenamefont {Chinn}\ \emph {et~al.}(1993)\citenamefont {Chinn},
  \citenamefont {Elster},\ and\ \citenamefont {Thaler}}]{chin93}%
  \BibitemOpen
  \bibfield  {author} {\bibinfo {author} {\bibfnamefont {C.~R.}\ \bibnamefont
  {Chinn}}, \bibinfo {author} {\bibfnamefont {C.}~\bibnamefont {Elster}}, \
  and\ \bibinfo {author} {\bibfnamefont {R.~M.}\ \bibnamefont {Thaler}},\
  }\href@noop {} {\bibfield  {journal} {\bibinfo  {journal} {Phys. Rev. C}\
  }\textbf {\bibinfo {volume} {48}},\ \bibinfo {pages} {2956} (\bibinfo {year}
  {1993})}\BibitemShut {NoStop}%
\bibitem [{\citenamefont {Crespo}\ \emph {et~al.}(1993)\citenamefont {Crespo},
  \citenamefont {Johnson},\ and\ \citenamefont {Tostevin}}]{crespo93}%
  \BibitemOpen
  \bibfield  {author} {\bibinfo {author} {\bibfnamefont {R.}~\bibnamefont
  {Crespo}}, \bibinfo {author} {\bibfnamefont {R.~C.}\ \bibnamefont {Johnson}},
  \ and\ \bibinfo {author} {\bibfnamefont {J.~A.}\ \bibnamefont {Tostevin}},\
  }\href@noop {} {\bibfield  {journal} {\bibinfo  {journal} {Phys. Rev. C}\
  }\textbf {\bibinfo {volume} {48}},\ \bibinfo {pages} {351} (\bibinfo {year}
  {1993})}\BibitemShut {NoStop}%
\bibitem [{\citenamefont {Waldecker}\ \emph {et~al.}(2011)\citenamefont
  {Waldecker}, \citenamefont {Barbieri},\ and\ \citenamefont
  {Dickhoff}}]{waldecker11}%
  \BibitemOpen
  \bibfield  {author} {\bibinfo {author} {\bibfnamefont {S.~J.}\ \bibnamefont
  {Waldecker}}, \bibinfo {author} {\bibfnamefont {C.}~\bibnamefont {Barbieri}},
  \ and\ \bibinfo {author} {\bibfnamefont {W.~H.}\ \bibnamefont {Dickhoff}},\
  }\href@noop {} {\bibfield  {journal} {\bibinfo  {journal} {Phys. Rev. C}\
  }\textbf {\bibinfo {volume} {84}},\ \bibinfo {pages} {034616} (\bibinfo
  {year} {2011})}\BibitemShut {NoStop}%
\bibitem [{\citenamefont {Charity}\ \emph {et~al.}(2007)\citenamefont
  {Charity}, \citenamefont {Mueller}, \citenamefont {Sobotka},\ and\
  \citenamefont {Dickhoff}}]{Dickhoff07}%
  \BibitemOpen
  \bibfield  {author} {\bibinfo {author} {\bibfnamefont {R.~J.}\ \bibnamefont
  {Charity}}, \bibinfo {author} {\bibfnamefont {J.~M.}\ \bibnamefont
  {Mueller}}, \bibinfo {author} {\bibfnamefont {L.~G.}\ \bibnamefont
  {Sobotka}}, \ and\ \bibinfo {author} {\bibfnamefont {W.~H.}\ \bibnamefont
  {Dickhoff}},\ }\href {\doibase 10.1103/PhysRevC.76.044314} {\bibfield
  {journal} {\bibinfo  {journal} {Phys. Rev. C}\ }\textbf {\bibinfo {volume}
  {76}},\ \bibinfo {pages} {044314} (\bibinfo {year} {2007})}\BibitemShut
  {NoStop}%
\bibitem [{\citenamefont {{Egashira}}\ \emph {et~al.}(2014)\citenamefont
  {{Egashira}}, \citenamefont {{Minomo}}, \citenamefont {{Toyokawa}},
  \citenamefont {{Matsumoto}},\ and\ \citenamefont {{Yahiro}}}]{Egashira}%
  \BibitemOpen
  \bibfield  {author} {\bibinfo {author} {\bibfnamefont {K.}~\bibnamefont
  {{Egashira}}}, \bibinfo {author} {\bibfnamefont {K.}~\bibnamefont
  {{Minomo}}}, \bibinfo {author} {\bibfnamefont {M.}~\bibnamefont
  {{Toyokawa}}}, \bibinfo {author} {\bibfnamefont {T.}~\bibnamefont
  {{Matsumoto}}}, \ and\ \bibinfo {author} {\bibfnamefont {M.}~\bibnamefont
  {{Yahiro}}},\ }\href {\doibase 10.1103/PhysRevC.89.064611} {\bibfield
  {journal} {\bibinfo  {journal} {\prc}\ }\textbf {\bibinfo {volume} {89}},\
  \bibinfo {eid} {064611} (\bibinfo {year} {2014})}\BibitemShut {NoStop}%
\bibitem [{\citenamefont {Rotureau}\ \emph {et~al.}(2018)\citenamefont
  {Rotureau}, \citenamefont {Danielewicz}, \citenamefont {Hagen}, \citenamefont
  {Jansen},\ and\ \citenamefont {Nunes}}]{Rotureau18}%
  \BibitemOpen
  \bibfield  {author} {\bibinfo {author} {\bibfnamefont {J.}~\bibnamefont
  {Rotureau}}, \bibinfo {author} {\bibfnamefont {P.}~\bibnamefont
  {Danielewicz}}, \bibinfo {author} {\bibfnamefont {G.}~\bibnamefont {Hagen}},
  \bibinfo {author} {\bibfnamefont {G.~R.}\ \bibnamefont {Jansen}}, \ and\
  \bibinfo {author} {\bibfnamefont {F.~M.}\ \bibnamefont {Nunes}},\ }\href@noop
  {} {\  (\bibinfo {year} {2018})},\ \Eprint {http://arxiv.org/abs/1808.04535}
  {arXiv:1808.04535 [nucl-th]} \BibitemShut {NoStop}%
\bibitem [{\citenamefont {Rotureau}\ \emph {et~al.}(2017)\citenamefont
  {Rotureau}, \citenamefont {Danielewicz}, \citenamefont {Hagen}, \citenamefont
  {Nunes},\ and\ \citenamefont {Papenbrock}}]{Rotureau17}%
  \BibitemOpen
  \bibfield  {author} {\bibinfo {author} {\bibfnamefont {J.}~\bibnamefont
  {Rotureau}}, \bibinfo {author} {\bibfnamefont {P.}~\bibnamefont
  {Danielewicz}}, \bibinfo {author} {\bibfnamefont {G.}~\bibnamefont {Hagen}},
  \bibinfo {author} {\bibfnamefont {F.~M.}\ \bibnamefont {Nunes}}, \ and\
  \bibinfo {author} {\bibfnamefont {T.}~\bibnamefont {Papenbrock}},\ }\href
  {\doibase 10.1103/PhysRevC.95.024315} {\bibfield  {journal} {\bibinfo
  {journal} {Phys. Rev. C}\ }\textbf {\bibinfo {volume} {95}},\ \bibinfo
  {pages} {024315} (\bibinfo {year} {2017})}\BibitemShut {NoStop}%
\bibitem [{\citenamefont {Toyokawa}\ \emph
  {et~al.}(2015{\natexlab{b}})\citenamefont {Toyokawa}, \citenamefont {Yahiro},
  \citenamefont {Matsumoto}, \citenamefont {Minomo}, \citenamefont {Ogata},\
  and\ \citenamefont {Kohno}}]{Toyokawa15}%
  \BibitemOpen
  \bibfield  {author} {\bibinfo {author} {\bibfnamefont {M.}~\bibnamefont
  {Toyokawa}}, \bibinfo {author} {\bibfnamefont {M.}~\bibnamefont {Yahiro}},
  \bibinfo {author} {\bibfnamefont {T.}~\bibnamefont {Matsumoto}}, \bibinfo
  {author} {\bibfnamefont {K.}~\bibnamefont {Minomo}}, \bibinfo {author}
  {\bibfnamefont {K.}~\bibnamefont {Ogata}}, \ and\ \bibinfo {author}
  {\bibfnamefont {M.}~\bibnamefont {Kohno}},\ }\href {\doibase
  10.1103/PhysRevC.92.024618} {\bibfield  {journal} {\bibinfo  {journal} {Phys.
  Rev. C}\ }\textbf {\bibinfo {volume} {92}},\ \bibinfo {pages} {024618}
  (\bibinfo {year} {2015}{\natexlab{b}})}\BibitemShut {NoStop}%
\bibitem [{\citenamefont {Durant}\ \emph {et~al.}(2018)\citenamefont {Durant},
  \citenamefont {Capel}, \citenamefont {Huth}, \citenamefont {Balantekin},\
  and\ \citenamefont {Schwenk}}]{DURANT18}%
  \BibitemOpen
  \bibfield  {author} {\bibinfo {author} {\bibfnamefont {V.}~\bibnamefont
  {Durant}}, \bibinfo {author} {\bibfnamefont {P.}~\bibnamefont {Capel}},
  \bibinfo {author} {\bibfnamefont {L.}~\bibnamefont {Huth}}, \bibinfo {author}
  {\bibfnamefont {A.}~\bibnamefont {Balantekin}}, \ and\ \bibinfo {author}
  {\bibfnamefont {A.}~\bibnamefont {Schwenk}},\ }\href {\doibase
  https://doi.org/10.1016/j.physletb.2018.05.084} {\bibfield  {journal}
  {\bibinfo  {journal} {Phys. Lett.}\ }\textbf {\bibinfo {volume} {B782}},\
  \bibinfo {pages} {668 } (\bibinfo {year} {2018})}\BibitemShut {NoStop}%
\bibitem [{\citenamefont {Weinberg}(1979)}]{WEINBERG79}%
  \BibitemOpen
  \bibfield  {author} {\bibinfo {author} {\bibfnamefont {S.}~\bibnamefont
  {Weinberg}},\ }\href {\doibase https://doi.org/10.1016/0378-4371(79)90223-1}
  {\bibfield  {journal} {\bibinfo  {journal} {Physica A}\ }\textbf {\bibinfo
  {volume} {96}},\ \bibinfo {pages} {327 } (\bibinfo {year}
  {1979})}\BibitemShut {NoStop}%
\bibitem [{\citenamefont {Epelbaum}\ \emph {et~al.}(2009)\citenamefont
  {Epelbaum}, \citenamefont {Hammer},\ and\ \citenamefont
  {Mei\ss{}ner}}]{epelbaum09}%
  \BibitemOpen
  \bibfield  {author} {\bibinfo {author} {\bibfnamefont {E.}~\bibnamefont
  {Epelbaum}}, \bibinfo {author} {\bibfnamefont {H.-W.}\ \bibnamefont
  {Hammer}}, \ and\ \bibinfo {author} {\bibfnamefont {U.-G.}\ \bibnamefont
  {Mei\ss{}ner}},\ }\href@noop {} {\bibfield  {journal} {\bibinfo  {journal}
  {Rev. Mod. Phys.}\ }\textbf {\bibinfo {volume} {81}},\ \bibinfo {pages}
  {1773} (\bibinfo {year} {2009})}\BibitemShut {NoStop}%
\bibitem [{\citenamefont {Machleidt}\ and\ \citenamefont
  {Entem}(2011)}]{MACHLEIDT11}%
  \BibitemOpen
  \bibfield  {author} {\bibinfo {author} {\bibfnamefont {R.}~\bibnamefont
  {Machleidt}}\ and\ \bibinfo {author} {\bibfnamefont {D.~R.}\ \bibnamefont
  {Entem}},\ }\href {\doibase https://doi.org/10.1016/j.physrep.2011.02.001}
  {\bibfield  {journal} {\bibinfo  {journal} {Phys. Rep.}\ }\textbf {\bibinfo
  {volume} {503}},\ \bibinfo {pages} {1 } (\bibinfo {year} {2011})}\BibitemShut
  {NoStop}%
\bibitem [{\citenamefont {Koning}\ \emph {et~al.}(2008)\citenamefont {Koning},
  \citenamefont {Hilaire},\ and\ \citenamefont {Duijvestijn}}]{TALYS}%
  \BibitemOpen
  \bibfield  {author} {\bibinfo {author} {\bibfnamefont {A.~J.}\ \bibnamefont
  {Koning}}, \bibinfo {author} {\bibfnamefont {S.}~\bibnamefont {Hilaire}}, \
  and\ \bibinfo {author} {\bibfnamefont {M.~C.}\ \bibnamefont {Duijvestijn}},\
  }\href@noop {} {\bibfield  {journal} {\bibinfo  {journal} {Proc. of the Int.
  Conf. on Nucl. Data for Science and Technology, EDP Sciences}\ } (\bibinfo
  {year} {2008})}\BibitemShut {NoStop}%
\bibitem [{\citenamefont {Tews}\ \emph {et~al.}(2013)\citenamefont {Tews},
  \citenamefont {Kr{\"u}ger}, \citenamefont {Hebeler},\ and\ \citenamefont
  {Schwenk}}]{tews13}%
  \BibitemOpen
  \bibfield  {author} {\bibinfo {author} {\bibfnamefont {I.}~\bibnamefont
  {Tews}}, \bibinfo {author} {\bibfnamefont {T.}~\bibnamefont {Kr{\"u}ger}},
  \bibinfo {author} {\bibfnamefont {K.}~\bibnamefont {Hebeler}}, \ and\
  \bibinfo {author} {\bibfnamefont {A.}~\bibnamefont {Schwenk}},\ }\href@noop
  {} {\bibfield  {journal} {\bibinfo  {journal} {Phys. Rev. Lett.}\ }\textbf
  {\bibinfo {volume} {110}},\ \bibinfo {pages} {032504} (\bibinfo {year}
  {2013})}\BibitemShut {NoStop}%
\bibitem [{\citenamefont {Drischler}\ \emph {et~al.}(2016)\citenamefont
  {Drischler}, \citenamefont {Carbone}, \citenamefont {Hebeler},\ and\
  \citenamefont {Schwenk}}]{drischler16}%
  \BibitemOpen
  \bibfield  {author} {\bibinfo {author} {\bibfnamefont {C.}~\bibnamefont
  {Drischler}}, \bibinfo {author} {\bibfnamefont {A.}~\bibnamefont {Carbone}},
  \bibinfo {author} {\bibfnamefont {K.}~\bibnamefont {Hebeler}}, \ and\
  \bibinfo {author} {\bibfnamefont {A.}~\bibnamefont {Schwenk}},\ }\href@noop
  {} {\bibfield  {journal} {\bibinfo  {journal} {Phys. Rev. C}\ }\textbf
  {\bibinfo {volume} {94}},\ \bibinfo {pages} {054307} (\bibinfo {year}
  {2016})}\BibitemShut {NoStop}%
\bibitem [{\citenamefont {Kaiser}\ and\ \citenamefont
  {Niessner}(2018)}]{kaiser18}%
  \BibitemOpen
  \bibfield  {author} {\bibinfo {author} {\bibfnamefont {N.}~\bibnamefont
  {Kaiser}}\ and\ \bibinfo {author} {\bibfnamefont {V.}~\bibnamefont
  {Niessner}},\ }\href {\doibase 10.1103/PhysRevC.98.054002} {\bibfield
  {journal} {\bibinfo  {journal} {Phys. Rev. C}\ }\textbf {\bibinfo {volume}
  {98}},\ \bibinfo {pages} {054002} (\bibinfo {year} {2018})}\BibitemShut
  {NoStop}%
\bibitem [{\citenamefont {Coraggio}\ \emph {et~al.}(2014)\citenamefont
  {Coraggio}, \citenamefont {Holt}, \citenamefont {Itaco}, \citenamefont
  {Machleidt}, \citenamefont {Marcucci},\ and\ \citenamefont
  {Sammarruca}}]{coraggio14}%
  \BibitemOpen
  \bibfield  {author} {\bibinfo {author} {\bibfnamefont {L.}~\bibnamefont
  {Coraggio}}, \bibinfo {author} {\bibfnamefont {J.~W.}\ \bibnamefont {Holt}},
  \bibinfo {author} {\bibfnamefont {N.}~\bibnamefont {Itaco}}, \bibinfo
  {author} {\bibfnamefont {R.}~\bibnamefont {Machleidt}}, \bibinfo {author}
  {\bibfnamefont {L.~E.}\ \bibnamefont {Marcucci}}, \ and\ \bibinfo {author}
  {\bibfnamefont {F.}~\bibnamefont {Sammarruca}},\ }\href@noop {} {\bibfield
  {journal} {\bibinfo  {journal} {Phys. Rev. C}\ }\textbf {\bibinfo {volume}
  {89}},\ \bibinfo {pages} {044321} (\bibinfo {year} {2014})}\BibitemShut
  {NoStop}%
\bibitem [{\citenamefont {Wellenhofer}\ \emph {et~al.}(2014)\citenamefont
  {Wellenhofer}, \citenamefont {Holt}, \citenamefont {Kaiser},\ and\
  \citenamefont {Weise}}]{wellenhofer14}%
  \BibitemOpen
  \bibfield  {author} {\bibinfo {author} {\bibfnamefont {C.}~\bibnamefont
  {Wellenhofer}}, \bibinfo {author} {\bibfnamefont {J.~W.}\ \bibnamefont
  {Holt}}, \bibinfo {author} {\bibfnamefont {N.}~\bibnamefont {Kaiser}}, \ and\
  \bibinfo {author} {\bibfnamefont {W.}~\bibnamefont {Weise}},\ }\href@noop {}
  {\bibfield  {journal} {\bibinfo  {journal} {Phys. Rev. C}\ }\textbf {\bibinfo
  {volume} {89}},\ \bibinfo {pages} {064009} (\bibinfo {year}
  {2014})}\BibitemShut {NoStop}%
\bibitem [{\citenamefont {Wellenhofer}\ \emph {et~al.}(2015)\citenamefont
  {Wellenhofer}, \citenamefont {Holt},\ and\ \citenamefont
  {Kaiser}}]{wellenhofer15}%
  \BibitemOpen
  \bibfield  {author} {\bibinfo {author} {\bibfnamefont {C.}~\bibnamefont
  {Wellenhofer}}, \bibinfo {author} {\bibfnamefont {J.~W.}\ \bibnamefont
  {Holt}}, \ and\ \bibinfo {author} {\bibfnamefont {N.}~\bibnamefont
  {Kaiser}},\ }\href@noop {} {\bibfield  {journal} {\bibinfo  {journal} {Phys.
  Rev. C}\ }\textbf {\bibinfo {volume} {92}},\ \bibinfo {pages} {015801}
  (\bibinfo {year} {2015})}\BibitemShut {NoStop}%
\bibitem [{\citenamefont {Holt}\ \emph {et~al.}(2018)\citenamefont {Holt},
  \citenamefont {Kaiser},\ and\ \citenamefont {Whitehead}}]{Holt18}%
  \BibitemOpen
  \bibfield  {author} {\bibinfo {author} {\bibfnamefont {J.~W.}\ \bibnamefont
  {Holt}}, \bibinfo {author} {\bibfnamefont {N.}~\bibnamefont {Kaiser}}, \ and\
  \bibinfo {author} {\bibfnamefont {T.~R.}\ \bibnamefont {Whitehead}},\ }\href
  {\doibase 10.1103/PhysRevC.97.054325} {\bibfield  {journal} {\bibinfo
  {journal} {Phys. Rev. C}\ }\textbf {\bibinfo {volume} {97}},\ \bibinfo
  {pages} {054325} (\bibinfo {year} {2018})}\BibitemShut {NoStop}%
\bibitem [{\citenamefont {Bell}\ and\ \citenamefont
  {Squires}(1959)}]{PhysRevLett.3.96}%
  \BibitemOpen
  \bibfield  {author} {\bibinfo {author} {\bibfnamefont {J.~S.}\ \bibnamefont
  {Bell}}\ and\ \bibinfo {author} {\bibfnamefont {E.~J.}\ \bibnamefont
  {Squires}},\ }\href {\doibase 10.1103/PhysRevLett.3.96} {\bibfield  {journal}
  {\bibinfo  {journal} {Phys. Rev. Lett.}\ }\textbf {\bibinfo {volume} {3}},\
  \bibinfo {pages} {96} (\bibinfo {year} {1959})}\BibitemShut {NoStop}%
\bibitem [{\citenamefont {Lim}\ and\ \citenamefont {Holt}(2017)}]{Lim17}%
  \BibitemOpen
  \bibfield  {author} {\bibinfo {author} {\bibfnamefont {Y.}~\bibnamefont
  {Lim}}\ and\ \bibinfo {author} {\bibfnamefont {J.~W.}\ \bibnamefont {Holt}},\
  }\href {\doibase 10.1103/PhysRevC.95.065805} {\bibfield  {journal} {\bibinfo
  {journal} {Phys. Rev. C}\ }\textbf {\bibinfo {volume} {95}},\ \bibinfo
  {pages} {065805} (\bibinfo {year} {2017})}\BibitemShut {NoStop}%
\bibitem [{\citenamefont {Jeukenne}\ \emph {et~al.}(1977)\citenamefont
  {Jeukenne}, \citenamefont {Lejeune},\ and\ \citenamefont
  {Mahaux}}]{Jeukenne77lda}%
  \BibitemOpen
  \bibfield  {author} {\bibinfo {author} {\bibfnamefont {J.~P.}\ \bibnamefont
  {Jeukenne}}, \bibinfo {author} {\bibfnamefont {A.}~\bibnamefont {Lejeune}}, \
  and\ \bibinfo {author} {\bibfnamefont {C.}~\bibnamefont {Mahaux}},\ }\href
  {\doibase 10.1103/PhysRevC.16.80} {\bibfield  {journal} {\bibinfo  {journal}
  {Phys. Rev. C}\ }\textbf {\bibinfo {volume} {16}},\ \bibinfo {pages} {80}
  (\bibinfo {year} {1977})}\BibitemShut {NoStop}%
\bibitem [{\citenamefont {Kosugi}\ and\ \citenamefont
  {Yoshida}(1982)}]{Kosugi82}%
  \BibitemOpen
  \bibfield  {author} {\bibinfo {author} {\bibfnamefont {S.}~\bibnamefont
  {Kosugi}}\ and\ \bibinfo {author} {\bibfnamefont {H.}~\bibnamefont
  {Yoshida}},\ }\href {\doibase 10.1016/0375-9474(82)90539-5} {\bibfield
  {journal} {\bibinfo  {journal} {Nucl. Phys.}\ }\textbf {\bibinfo {volume}
  {A373}},\ \bibinfo {pages} {349} (\bibinfo {year} {1982})}\BibitemShut
  {NoStop}%
\bibitem [{\citenamefont {Bogner}\ \emph {et~al.}(2009)\citenamefont {Bogner},
  \citenamefont {Furnstahl},\ and\ \citenamefont {Platter}}]{bogner08}%
  \BibitemOpen
  \bibfield  {author} {\bibinfo {author} {\bibfnamefont {S.~K.}\ \bibnamefont
  {Bogner}}, \bibinfo {author} {\bibfnamefont {R.~J.}\ \bibnamefont
  {Furnstahl}}, \ and\ \bibinfo {author} {\bibfnamefont {L.}~\bibnamefont
  {Platter}},\ }\href {\doibase 10.1140/epja/i2008-10695-1} {\bibfield
  {journal} {\bibinfo  {journal} {Eur. Phys. J. A}\ }\textbf {\bibinfo {volume}
  {39}},\ \bibinfo {pages} {219} (\bibinfo {year} {2009})}\BibitemShut
  {NoStop}%
\bibitem [{\citenamefont {Gebremariam}\ \emph {et~al.}(2010)\citenamefont
  {Gebremariam}, \citenamefont {Duguet},\ and\ \citenamefont
  {Bogner}}]{gebremariam10}%
  \BibitemOpen
  \bibfield  {author} {\bibinfo {author} {\bibfnamefont {B.}~\bibnamefont
  {Gebremariam}}, \bibinfo {author} {\bibfnamefont {T.}~\bibnamefont {Duguet}},
  \ and\ \bibinfo {author} {\bibfnamefont {S.~K.}\ \bibnamefont {Bogner}},\
  }\href@noop {} {\bibfield  {journal} {\bibinfo  {journal} {Phys. Rev. C}\
  }\textbf {\bibinfo {volume} {82}},\ \bibinfo {pages} {014305} (\bibinfo
  {year} {2010})}\BibitemShut {NoStop}%
\bibitem [{\citenamefont {Holt}\ \emph {et~al.}(2011)\citenamefont {Holt},
  \citenamefont {Kaiser},\ and\ \citenamefont {Weise}}]{KaiserHoltEDF}%
  \BibitemOpen
  \bibfield  {author} {\bibinfo {author} {\bibfnamefont {J.~W.}\ \bibnamefont
  {Holt}}, \bibinfo {author} {\bibfnamefont {N.}~\bibnamefont {Kaiser}}, \ and\
  \bibinfo {author} {\bibfnamefont {W.}~\bibnamefont {Weise}},\ }\href
  {https://doi.org/10.1140/epja/i2011-11128-x} {\bibfield  {journal} {\bibinfo
  {journal} {Eur. Phys. J. A}\ }\textbf {\bibinfo {volume} {47}} (\bibinfo
  {year} {2011})}\BibitemShut {NoStop}%
\bibitem [{\citenamefont {Negele}\ and\ \citenamefont
  {Vautherin}(1972)}]{negele72}%
  \BibitemOpen
  \bibfield  {author} {\bibinfo {author} {\bibfnamefont {J.~W.}\ \bibnamefont
  {Negele}}\ and\ \bibinfo {author} {\bibfnamefont {D.}~\bibnamefont
  {Vautherin}},\ }\href@noop {} {\bibfield  {journal} {\bibinfo  {journal}
  {Phys. Rev. C}\ }\textbf {\bibinfo {volume} {5}},\ \bibinfo {pages} {1472}
  (\bibinfo {year} {1972})}\BibitemShut {NoStop}%
\bibitem [{\citenamefont {Bogner}\ \emph {et~al.}(2005)\citenamefont {Bogner},
  \citenamefont {Schwenk}, \citenamefont {Furnstahl},\ and\ \citenamefont
  {Nogga}}]{bogner05}%
  \BibitemOpen
  \bibfield  {author} {\bibinfo {author} {\bibfnamefont {S.~K.}\ \bibnamefont
  {Bogner}}, \bibinfo {author} {\bibfnamefont {A.}~\bibnamefont {Schwenk}},
  \bibinfo {author} {\bibfnamefont {R.~J.}\ \bibnamefont {Furnstahl}}, \ and\
  \bibinfo {author} {\bibfnamefont {A.}~\bibnamefont {Nogga}},\ }\href@noop {}
  {\bibfield  {journal} {\bibinfo  {journal} {Nucl. Phys.}\ }\textbf {\bibinfo
  {volume} {A763}},\ \bibinfo {pages} {59} (\bibinfo {year}
  {2005})}\BibitemShut {NoStop}%
\bibitem [{\citenamefont {Holt}\ \emph {et~al.}(2009)\citenamefont {Holt},
  \citenamefont {Kaiser},\ and\ \citenamefont {Weise}}]{Holt09}%
  \BibitemOpen
  \bibfield  {author} {\bibinfo {author} {\bibfnamefont {J.~W.}\ \bibnamefont
  {Holt}}, \bibinfo {author} {\bibfnamefont {N.}~\bibnamefont {Kaiser}}, \ and\
  \bibinfo {author} {\bibfnamefont {W.}~\bibnamefont {Weise}},\ }\href
  {\doibase 10.1103/PhysRevC.79.054331} {\bibfield  {journal} {\bibinfo
  {journal} {Phys. Rev. C}\ }\textbf {\bibinfo {volume} {79}},\ \bibinfo
  {pages} {054331} (\bibinfo {year} {2009})}\BibitemShut {NoStop}%
\bibitem [{\citenamefont {Hebeler}\ and\ \citenamefont
  {Schwenk}(2010)}]{hebeler10}%
  \BibitemOpen
  \bibfield  {author} {\bibinfo {author} {\bibfnamefont {K.}~\bibnamefont
  {Hebeler}}\ and\ \bibinfo {author} {\bibfnamefont {A.}~\bibnamefont
  {Schwenk}},\ }\href@noop {} {\bibfield  {journal} {\bibinfo  {journal} {Phys.
  Rev. C}\ }\textbf {\bibinfo {volume} {82}},\ \bibinfo {pages} {014314}
  (\bibinfo {year} {2010})}\BibitemShut {NoStop}%
\bibitem [{\citenamefont {Negele}\ and\ \citenamefont
  {Yazaki}(1981)}]{Negele81}%
  \BibitemOpen
  \bibfield  {author} {\bibinfo {author} {\bibfnamefont {J.~W.}\ \bibnamefont
  {Negele}}\ and\ \bibinfo {author} {\bibfnamefont {K.}~\bibnamefont
  {Yazaki}},\ }\href {\doibase 10.1103/PhysRevLett.47.71} {\bibfield  {journal}
  {\bibinfo  {journal} {Phys. Rev. Lett.}\ }\textbf {\bibinfo {volume} {47}},\
  \bibinfo {pages} {71} (\bibinfo {year} {1981})}\BibitemShut {NoStop}%
\bibitem [{\citenamefont {Fantoni}\ \emph {et~al.}(1981)\citenamefont
  {Fantoni}, \citenamefont {Friman},\ and\ \citenamefont
  {Pandharipande}}]{fantoni81}%
  \BibitemOpen
  \bibfield  {author} {\bibinfo {author} {\bibfnamefont {S.}~\bibnamefont
  {Fantoni}}, \bibinfo {author} {\bibfnamefont {B.~L.}\ \bibnamefont {Friman}},
  \ and\ \bibinfo {author} {\bibfnamefont {V.~R.}\ \bibnamefont
  {Pandharipande}},\ }\href {\doibase 10.1016/0370-2693(81)90565-7} {\bibfield
  {journal} {\bibinfo  {journal} {Phys. Lett.}\ }\textbf {\bibinfo {volume}
  {B104}},\ \bibinfo {pages} {89} (\bibinfo {year} {1981})}\BibitemShut
  {NoStop}%
\bibitem [{\citenamefont {Gebremariam}\ \emph {et~al.}(2011)\citenamefont
  {Gebremariam}, \citenamefont {Bogner},\ and\ \citenamefont
  {Duguet}}]{Gebremariam10npa}%
  \BibitemOpen
  \bibfield  {author} {\bibinfo {author} {\bibfnamefont {B.}~\bibnamefont
  {Gebremariam}}, \bibinfo {author} {\bibfnamefont {S.~K.}\ \bibnamefont
  {Bogner}}, \ and\ \bibinfo {author} {\bibfnamefont {T.}~\bibnamefont
  {Duguet}},\ }\href@noop {} {\bibfield  {journal} {\bibinfo  {journal} {Nucl.
  Phys.}\ }\textbf {\bibinfo {volume} {A851}},\ \bibinfo {pages} {17} (\bibinfo
  {year} {2011})}\BibitemShut {NoStop}%
\bibitem [{\citenamefont {Zhang}\ \emph {et~al.}(2018)\citenamefont {Zhang},
  \citenamefont {Bogner},\ and\ \citenamefont {Furnstahl}}]{zhang18}%
  \BibitemOpen
  \bibfield  {author} {\bibinfo {author} {\bibfnamefont {Y.~N.}\ \bibnamefont
  {Zhang}}, \bibinfo {author} {\bibfnamefont {S.~K.}\ \bibnamefont {Bogner}}, \
  and\ \bibinfo {author} {\bibfnamefont {R.~J.}\ \bibnamefont {Furnstahl}},\
  }\href@noop {} {\bibfield  {journal} {\bibinfo  {journal} {Phys. Rev. C}\
  }\textbf {\bibinfo {volume} {98}},\ \bibinfo {pages} {064306} (\bibinfo
  {year} {2018})}\BibitemShut {NoStop}%
\bibitem [{\citenamefont {Kaiser}(2012)}]{Kaiser12}%
  \BibitemOpen
  \bibfield  {author} {\bibinfo {author} {\bibfnamefont {N.}~\bibnamefont
  {Kaiser}},\ }\href@noop {} {\bibfield  {journal} {\bibinfo  {journal} {Eur.
  Phys. J. A}\ }\textbf {\bibinfo {volume} {48}},\ \bibinfo {pages} {36}
  (\bibinfo {year} {2012})}\BibitemShut {NoStop}%
\bibitem [{\citenamefont {Brieva}\ and\ \citenamefont
  {Rook}(1977{\natexlab{b}})}]{BRIEVA1977317}%
  \BibitemOpen
  \bibfield  {author} {\bibinfo {author} {\bibfnamefont {F.~A.}\ \bibnamefont
  {Brieva}}\ and\ \bibinfo {author} {\bibfnamefont {J.~R.}\ \bibnamefont
  {Rook}},\ }\href {\doibase https://doi.org/10.1016/0375-9474(77)90323-2}
  {\bibfield  {journal} {\bibinfo  {journal} {Nucl. Phys.}\ }\textbf {\bibinfo
  {volume} {A291}},\ \bibinfo {pages} {317 } (\bibinfo {year}
  {1977}{\natexlab{b}})}\BibitemShut {NoStop}%
\bibitem [{\citenamefont {Wiringa}\ \emph {et~al.}(1995)\citenamefont
  {Wiringa}, \citenamefont {Stoks},\ and\ \citenamefont
  {Schiavilla}}]{Argonne}%
  \BibitemOpen
  \bibfield  {author} {\bibinfo {author} {\bibfnamefont {R.~B.}\ \bibnamefont
  {Wiringa}}, \bibinfo {author} {\bibfnamefont {V.~G.~J.}\ \bibnamefont
  {Stoks}}, \ and\ \bibinfo {author} {\bibfnamefont {R.}~\bibnamefont
  {Schiavilla}},\ }\href {\doibase 10.1103/PhysRevC.51.38} {\bibfield
  {journal} {\bibinfo  {journal} {Phys. Rev. C}\ }\textbf {\bibinfo {volume}
  {51}},\ \bibinfo {pages} {38} (\bibinfo {year} {1995})}\BibitemShut {NoStop}%
\bibitem [{\citenamefont {Lagrange}\ and\ \citenamefont
  {Lejeune}(1982)}]{Lagrange82}%
  \BibitemOpen
  \bibfield  {author} {\bibinfo {author} {\bibfnamefont {C.}~\bibnamefont
  {Lagrange}}\ and\ \bibinfo {author} {\bibfnamefont {A.}~\bibnamefont
  {Lejeune}},\ }\href {\doibase 10.1103/PhysRevC.25.2278} {\bibfield  {journal}
  {\bibinfo  {journal} {Phys. Rev. C}\ }\textbf {\bibinfo {volume} {25}},\
  \bibinfo {pages} {2278} (\bibinfo {year} {1982})}\BibitemShut {NoStop}%
\bibitem [{\citenamefont {Kohno}\ \emph {et~al.}(1984)\citenamefont {Kohno},
  \citenamefont {Sprung}, \citenamefont {Nagata},\ and\ \citenamefont
  {Yamaguchi}}]{Kohno84}%
  \BibitemOpen
  \bibfield  {author} {\bibinfo {author} {\bibfnamefont {M.}~\bibnamefont
  {Kohno}}, \bibinfo {author} {\bibfnamefont {D.~W.~L.}\ \bibnamefont
  {Sprung}}, \bibinfo {author} {\bibfnamefont {S.}~\bibnamefont {Nagata}}, \
  and\ \bibinfo {author} {\bibfnamefont {N.}~\bibnamefont {Yamaguchi}},\ }\href
  {\doibase 10.1016/0370-2693(84)91095-5} {\bibfield  {journal} {\bibinfo
  {journal} {Phys. Lett.}\ }\textbf {\bibinfo {volume} {B137}},\ \bibinfo
  {pages} {10} (\bibinfo {year} {1984})}\BibitemShut {NoStop}%
\bibitem [{\citenamefont {Holt}\ \emph
  {et~al.}(2016{\natexlab{b}})\citenamefont {Holt}, \citenamefont {Rho},\ and\
  \citenamefont {Weise}}]{holt16pr}%
  \BibitemOpen
  \bibfield  {author} {\bibinfo {author} {\bibfnamefont {J.~W.}\ \bibnamefont
  {Holt}}, \bibinfo {author} {\bibfnamefont {M.}~\bibnamefont {Rho}}, \ and\
  \bibinfo {author} {\bibfnamefont {W.}~\bibnamefont {Weise}},\ }\href@noop {}
  {\bibfield  {journal} {\bibinfo  {journal} {Phys. Rept.}\ }\textbf {\bibinfo
  {volume} {621}},\ \bibinfo {pages} {2} (\bibinfo {year}
  {2016}{\natexlab{b}})}\BibitemShut {NoStop}%
\bibitem [{\citenamefont {Kaiser}\ and\ \citenamefont
  {Weise}(2010)}]{kaiser10}%
  \BibitemOpen
  \bibfield  {author} {\bibinfo {author} {\bibfnamefont {N.}~\bibnamefont
  {Kaiser}}\ and\ \bibinfo {author} {\bibfnamefont {W.}~\bibnamefont {Weise}},\
  }\href@noop {} {\bibfield  {journal} {\bibinfo  {journal} {Nucl. Phys.}\
  }\textbf {\bibinfo {volume} {A836}},\ \bibinfo {pages} {256} (\bibinfo {year}
  {2010})}\BibitemShut {NoStop}%
\bibitem [{\citenamefont {Brieva}\ and\ \citenamefont
  {Rook}(1978)}]{BRIEVA1978206}%
  \BibitemOpen
  \bibfield  {author} {\bibinfo {author} {\bibfnamefont {F.~A.}\ \bibnamefont
  {Brieva}}\ and\ \bibinfo {author} {\bibfnamefont {J.~R.}\ \bibnamefont
  {Rook}},\ }\href {\doibase https://doi.org/10.1016/0375-9474(78)90272-5}
  {\bibfield  {journal} {\bibinfo  {journal} {Nucl. Phys.}\ }\textbf {\bibinfo
  {volume} {A297}},\ \bibinfo {pages} {206 } (\bibinfo {year}
  {1978})}\BibitemShut {NoStop}%
\bibitem [{\citenamefont {McCamis}\ \emph {et~al.}(1986)\citenamefont
  {McCamis}, \citenamefont {Nasr}, \citenamefont {Birchall}, \citenamefont
  {Davison}, \citenamefont {van Oers}, \citenamefont {Verheijen}, \citenamefont
  {Carlson}, \citenamefont {Cox}, \citenamefont {Clark}, \citenamefont
  {Cooper}, \citenamefont {Hama},\ and\ \citenamefont {Mercer}}]{Calciumdata}%
  \BibitemOpen
  \bibfield  {author} {\bibinfo {author} {\bibfnamefont {R.~H.}\ \bibnamefont
  {McCamis}}, \bibinfo {author} {\bibfnamefont {T.~N.}\ \bibnamefont {Nasr}},
  \bibinfo {author} {\bibfnamefont {J.}~\bibnamefont {Birchall}}, \bibinfo
  {author} {\bibfnamefont {N.~E.}\ \bibnamefont {Davison}}, \bibinfo {author}
  {\bibfnamefont {W.~T.~H.}\ \bibnamefont {van Oers}}, \bibinfo {author}
  {\bibfnamefont {P.~J.~T.}\ \bibnamefont {Verheijen}}, \bibinfo {author}
  {\bibfnamefont {R.~F.}\ \bibnamefont {Carlson}}, \bibinfo {author}
  {\bibfnamefont {A.~J.}\ \bibnamefont {Cox}}, \bibinfo {author} {\bibfnamefont
  {B.~C.}\ \bibnamefont {Clark}}, \bibinfo {author} {\bibfnamefont {E.~D.}\
  \bibnamefont {Cooper}}, \bibinfo {author} {\bibfnamefont {S.}~\bibnamefont
  {Hama}}, \ and\ \bibinfo {author} {\bibfnamefont {R.~L.}\ \bibnamefont
  {Mercer}},\ }\href {\doibase 10.1103/PhysRevC.33.1624} {\bibfield  {journal}
  {\bibinfo  {journal} {Phys. Rev. C}\ }\textbf {\bibinfo {volume} {33}},\
  \bibinfo {pages} {1624} (\bibinfo {year} {1986})}\BibitemShut {NoStop}%
\bibitem [{\citenamefont {Yagi}\ \emph {et~al.}(1964)\citenamefont {Yagi},
  \citenamefont {Ejiri}, \citenamefont {Furukawa}, \citenamefont {Ishizaki},
  \citenamefont {Koike}, \citenamefont {Matsuda}, \citenamefont {Nakajima},
  \citenamefont {Nonaka}, \citenamefont {Saji}, \citenamefont {Tanaka},\ and\
  \citenamefont {Satchler}}]{Ca40e55}%
  \BibitemOpen
  \bibfield  {author} {\bibinfo {author} {\bibfnamefont {K.}~\bibnamefont
  {Yagi}}, \bibinfo {author} {\bibfnamefont {H.}~\bibnamefont {Ejiri}},
  \bibinfo {author} {\bibfnamefont {M.}~\bibnamefont {Furukawa}}, \bibinfo
  {author} {\bibfnamefont {Y.}~\bibnamefont {Ishizaki}}, \bibinfo {author}
  {\bibfnamefont {M.}~\bibnamefont {Koike}}, \bibinfo {author} {\bibfnamefont
  {K.}~\bibnamefont {Matsuda}}, \bibinfo {author} {\bibfnamefont
  {Y.}~\bibnamefont {Nakajima}}, \bibinfo {author} {\bibfnamefont
  {I.}~\bibnamefont {Nonaka}}, \bibinfo {author} {\bibfnamefont
  {Y.}~\bibnamefont {Saji}}, \bibinfo {author} {\bibfnamefont {E.}~\bibnamefont
  {Tanaka}}, \ and\ \bibinfo {author} {\bibfnamefont {G.}~\bibnamefont
  {Satchler}},\ }\href {\doibase https://doi.org/10.1016/0031-9163(64)90164-7}
  {\bibfield  {journal} {\bibinfo  {journal} {Phys. Lett.}\ }\textbf {\bibinfo
  {volume} {10}},\ \bibinfo {pages} {186 } (\bibinfo {year}
  {1964})}\BibitemShut {NoStop}%
\bibitem [{\citenamefont {Sakaguchi}\ \emph {et~al.}(1979)\citenamefont
  {Sakaguchi}, \citenamefont {Nakamura}, \citenamefont {Hatanaka},
  \citenamefont {Goto}, \citenamefont {Noro}, \citenamefont {Ohtani},
  \citenamefont {Sakamoto},\ and\ \citenamefont {Kobayashi}}]{Ca4065}%
  \BibitemOpen
  \bibfield  {author} {\bibinfo {author} {\bibfnamefont {H.}~\bibnamefont
  {Sakaguchi}}, \bibinfo {author} {\bibfnamefont {M.}~\bibnamefont {Nakamura}},
  \bibinfo {author} {\bibfnamefont {K.}~\bibnamefont {Hatanaka}}, \bibinfo
  {author} {\bibfnamefont {A.}~\bibnamefont {Goto}}, \bibinfo {author}
  {\bibfnamefont {T.}~\bibnamefont {Noro}}, \bibinfo {author} {\bibfnamefont
  {F.}~\bibnamefont {Ohtani}}, \bibinfo {author} {\bibfnamefont
  {H.}~\bibnamefont {Sakamoto}}, \ and\ \bibinfo {author} {\bibfnamefont
  {S.}~\bibnamefont {Kobayashi}},\ }\href {\doibase
  https://doi.org/10.1016/0370-2693(79)90071-6} {\bibfield  {journal} {\bibinfo
   {journal} {Phys. Lett.}\ }\textbf {\bibinfo {volume} {B89}},\ \bibinfo
  {pages} {40 } (\bibinfo {year} {1979})}\BibitemShut {NoStop}%
\bibitem [{\citenamefont {Nadasen}\ \emph {et~al.}(1981)\citenamefont
  {Nadasen}, \citenamefont {Schwandt}, \citenamefont {Singh}, \citenamefont
  {Jacobs}, \citenamefont {Bacher}, \citenamefont {Debevec}, \citenamefont
  {Kaitchuck},\ and\ \citenamefont {Meek}}]{Ca40e80160}%
  \BibitemOpen
  \bibfield  {author} {\bibinfo {author} {\bibfnamefont {A.}~\bibnamefont
  {Nadasen}}, \bibinfo {author} {\bibfnamefont {P.}~\bibnamefont {Schwandt}},
  \bibinfo {author} {\bibfnamefont {P.~P.}\ \bibnamefont {Singh}}, \bibinfo
  {author} {\bibfnamefont {W.~W.}\ \bibnamefont {Jacobs}}, \bibinfo {author}
  {\bibfnamefont {A.~D.}\ \bibnamefont {Bacher}}, \bibinfo {author}
  {\bibfnamefont {P.~T.}\ \bibnamefont {Debevec}}, \bibinfo {author}
  {\bibfnamefont {M.~D.}\ \bibnamefont {Kaitchuck}}, \ and\ \bibinfo {author}
  {\bibfnamefont {J.~T.}\ \bibnamefont {Meek}},\ }\href {\doibase
  10.1103/PhysRevC.23.1023} {\bibfield  {journal} {\bibinfo  {journal} {Phys.
  Rev. C}\ }\textbf {\bibinfo {volume} {23}},\ \bibinfo {pages} {1023}
  (\bibinfo {year} {1981})}\BibitemShut {NoStop}%
\bibitem [{\citenamefont {Koltay}\ \emph {et~al.}(1975)\citenamefont {Koltay},
  \citenamefont {Mesk\'{o}},\ and\ \citenamefont {V\'{e}gh}}]{KOLTAY1975173}%
  \BibitemOpen
  \bibfield  {author} {\bibinfo {author} {\bibfnamefont {E.}~\bibnamefont
  {Koltay}}, \bibinfo {author} {\bibfnamefont {L.}~\bibnamefont {Mesk\'{o}}}, \
  and\ \bibinfo {author} {\bibfnamefont {L.}~\bibnamefont {V\'{e}gh}},\ }\href
  {\doibase https://doi.org/10.1016/0375-9474(75)90098-6} {\bibfield  {journal}
  {\bibinfo  {journal} {Nucl. Phys.}\ }\textbf {\bibinfo {volume} {A249}},\
  \bibinfo {pages} {173 } (\bibinfo {year} {1975})}\BibitemShut {NoStop}%
\bibitem [{\citenamefont {Auce}\ \emph {et~al.}(2005)\citenamefont {Auce},
  \citenamefont {Ingemarsson}, \citenamefont {Johansson}, \citenamefont
  {Lantz}, \citenamefont {Tibell}, \citenamefont {Carlson}, \citenamefont
  {Shachno}, \citenamefont {Cowley}, \citenamefont {Hillhouse}, \citenamefont
  {Jacobs}, \citenamefont {Stander}, \citenamefont {Zyl}, \citenamefont
  {F\"ortsch}, \citenamefont {Lawrie}, \citenamefont {Smit},\ and\
  \citenamefont {Steyn}}]{reactionCSCa40}%
  \BibitemOpen
  \bibfield  {author} {\bibinfo {author} {\bibfnamefont {A.}~\bibnamefont
  {Auce}}, \bibinfo {author} {\bibfnamefont {A.}~\bibnamefont {Ingemarsson}},
  \bibinfo {author} {\bibfnamefont {R.}~\bibnamefont {Johansson}}, \bibinfo
  {author} {\bibfnamefont {M.}~\bibnamefont {Lantz}}, \bibinfo {author}
  {\bibfnamefont {G.}~\bibnamefont {Tibell}}, \bibinfo {author} {\bibfnamefont
  {R.~F.}\ \bibnamefont {Carlson}}, \bibinfo {author} {\bibfnamefont {M.~J.}\
  \bibnamefont {Shachno}}, \bibinfo {author} {\bibfnamefont {A.~A.}\
  \bibnamefont {Cowley}}, \bibinfo {author} {\bibfnamefont {G.~C.}\
  \bibnamefont {Hillhouse}}, \bibinfo {author} {\bibfnamefont {N.~M.}\
  \bibnamefont {Jacobs}}, \bibinfo {author} {\bibfnamefont {J.~A.}\
  \bibnamefont {Stander}}, \bibinfo {author} {\bibfnamefont {J.~J.}\
  \bibnamefont {vanZyl}}, \bibinfo {author} {\bibfnamefont {S.~V.}\ \bibnamefont
  {F\"ortsch}}, \bibinfo {author} {\bibfnamefont {J.~J.}\ \bibnamefont
  {Lawrie}}, \bibinfo {author} {\bibfnamefont {F.~D.}\ \bibnamefont {Smit}}, \
  and\ \bibinfo {author} {\bibfnamefont {G.~F.}\ \bibnamefont {Steyn}},\ }\href
  {\doibase 10.1103/PhysRevC.71.064606} {\bibfield  {journal} {\bibinfo
  {journal} {Phys. Rev. C}\ }\textbf {\bibinfo {volume} {71}},\ \bibinfo
  {pages} {064606} (\bibinfo {year} {2005})}\BibitemShut {NoStop}%
\bibitem [{\citenamefont {Dicello}\ and\ \citenamefont
  {Igo}(1970)}]{reactionCSCa40low}%
  \BibitemOpen
  \bibfield  {author} {\bibinfo {author} {\bibfnamefont {J.~F.}\ \bibnamefont
  {Dicello}}\ and\ \bibinfo {author} {\bibfnamefont {G.}~\bibnamefont {Igo}},\
  }\href {\doibase 10.1103/PhysRevC.2.488} {\bibfield  {journal} {\bibinfo
  {journal} {Phys. Rev. C}\ }\textbf {\bibinfo {volume} {2}},\ \bibinfo {pages}
  {488} (\bibinfo {year} {1970})}\BibitemShut {NoStop}%
\bibitem [{\citenamefont {Carlson}\ \emph {et~al.}(1975)\citenamefont
  {Carlson}, \citenamefont {Cox}, \citenamefont {Nimmo}, \citenamefont
  {Davison}, \citenamefont {Elbakr}, \citenamefont {Horton}, \citenamefont
  {Houdayer}, \citenamefont {Sourkes}, \citenamefont {van Oers},\ and\
  \citenamefont {Margaziotis}}]{reactionCSCa40mid}%
  \BibitemOpen
  \bibfield  {author} {\bibinfo {author} {\bibfnamefont {R.~F.}\ \bibnamefont
  {Carlson}}, \bibinfo {author} {\bibfnamefont {A.~J.}\ \bibnamefont {Cox}},
  \bibinfo {author} {\bibfnamefont {J.~R.}\ \bibnamefont {Nimmo}}, \bibinfo
  {author} {\bibfnamefont {N.~E.}\ \bibnamefont {Davison}}, \bibinfo {author}
  {\bibfnamefont {S.~A.}\ \bibnamefont {Elbakr}}, \bibinfo {author}
  {\bibfnamefont {J.~L.}\ \bibnamefont {Horton}}, \bibinfo {author}
  {\bibfnamefont {A.}~\bibnamefont {Houdayer}}, \bibinfo {author}
  {\bibfnamefont {A.~M.}\ \bibnamefont {Sourkes}}, \bibinfo {author}
  {\bibfnamefont {W.~T.~H.}\ \bibnamefont {van Oers}}, \ and\ \bibinfo {author}
  {\bibfnamefont {D.~J.}\ \bibnamefont {Margaziotis}},\ }\href {\doibase
  10.1103/PhysRevC.12.1167} {\bibfield  {journal} {\bibinfo  {journal} {Phys.
  Rev. C}\ }\textbf {\bibinfo {volume} {12}},\ \bibinfo {pages} {1167}
  (\bibinfo {year} {1975})}\BibitemShut {NoStop}%
\bibitem [{\citenamefont {Carlson}\ \emph {et~al.}(1994)\citenamefont
  {Carlson}, \citenamefont {Cox}, \citenamefont {Davison}, \citenamefont
  {Eliyakut-Roshko}, \citenamefont {McCamis},\ and\ \citenamefont {van
  Oers}}]{reactionCSCa424448}%
  \BibitemOpen
  \bibfield  {author} {\bibinfo {author} {\bibfnamefont {R.~F.}\ \bibnamefont
  {Carlson}}, \bibinfo {author} {\bibfnamefont {A.~J.}\ \bibnamefont {Cox}},
  \bibinfo {author} {\bibfnamefont {N.~E.}\ \bibnamefont {Davison}}, \bibinfo
  {author} {\bibfnamefont {T.}~\bibnamefont {Eliyakut-Roshko}}, \bibinfo
  {author} {\bibfnamefont {R.~H.}\ \bibnamefont {McCamis}}, \ and\ \bibinfo
  {author} {\bibfnamefont {W.~T.~H.}\ \bibnamefont {van Oers}},\ }\href
  {\doibase 10.1103/PhysRevC.49.3090} {\bibfield  {journal} {\bibinfo
  {journal} {Phys. Rev. C}\ }\textbf {\bibinfo {volume} {49}},\ \bibinfo
  {pages} {3090} (\bibinfo {year} {1994})}\BibitemShut {NoStop}%
\bibitem [{\citenamefont {Mumpower}\ \emph {et~al.}(2016)\citenamefont
  {Mumpower}, \citenamefont {Surman}, \citenamefont {McLaughlin},\ and\
  \citenamefont {Aprahamian}}]{mumpower15}%
  \BibitemOpen
  \bibfield  {author} {\bibinfo {author} {\bibfnamefont {M.~R.}\ \bibnamefont
  {Mumpower}}, \bibinfo {author} {\bibfnamefont {R.}~\bibnamefont {Surman}},
  \bibinfo {author} {\bibfnamefont {G.~C.}\ \bibnamefont {McLaughlin}}, \ and\
  \bibinfo {author} {\bibfnamefont {A.}~\bibnamefont {Aprahamian}},\
  }\href@noop {} {\bibfield  {journal} {\bibinfo  {journal} {Prog. Part. Nucl.
  Phys.}\ }\textbf {\bibinfo {volume} {86}},\ \bibinfo {pages} {86} (\bibinfo
  {year} {2016})}\BibitemShut {NoStop}%
\bibitem [{\citenamefont {Horowitz}\ \emph {et~al.}(2018)\citenamefont
  {Horowitz} \emph {et~al.}}]{horowitz18}%
  \BibitemOpen
  \bibfield  {author} {\bibinfo {author} {\bibfnamefont {C.~J.}\ \bibnamefont
  {Horowitz}} \emph {et~al.},\ }\href@noop {} {\bibfield  {journal} {\bibinfo
  {journal} {arXiv:1805.04637}\ } (\bibinfo {year} {2018})}\BibitemShut
  {NoStop}%
\bibitem [{\citenamefont {Sammarruca}\ \emph {et~al.}(2018)\citenamefont
  {Sammarruca}, \citenamefont {Marcucci}, \citenamefont {Coraggio},
  \citenamefont {Holt}, \citenamefont {Itaco},\ and\ \citenamefont
  {Machleidt}}]{Sammarruca18}%
  \BibitemOpen
  \bibfield  {author} {\bibinfo {author} {\bibfnamefont {F.}~\bibnamefont
  {Sammarruca}}, \bibinfo {author} {\bibfnamefont {L.~E.}\ \bibnamefont
  {Marcucci}}, \bibinfo {author} {\bibfnamefont {L.}~\bibnamefont {Coraggio}},
  \bibinfo {author} {\bibfnamefont {J.~W.}\ \bibnamefont {Holt}}, \bibinfo
  {author} {\bibfnamefont {N.}~\bibnamefont {Itaco}}, \ and\ \bibinfo {author}
  {\bibfnamefont {R.}~\bibnamefont {Machleidt}},\ }\href@noop {} {\bibfield
  {journal} {\bibinfo  {journal} {arXiv:1807.06640}\ } (\bibinfo {year}
  {2018})}\BibitemShut {NoStop}%
\end{thebibliography}

%

\end{document}